\documentclass[a4paper,11pt]{article}

\usepackage{amsmath,amssymb,color,graphics,epsfig,cite}
\usepackage{titlesec}
\usepackage[titletoc,title]{appendix}

\usepackage{amsfonts}

\newcommand{\hoch}[1]{$\, ^{#1}$}

\newcommand{\be}{\begin{equation}}
	\newcommand{\ee}{\end{equation}}
\newcommand{\bea}{\setlength\arraycolsep{2pt} \begin{eqnarray}}
	\newcommand{\eea}{\end{eqnarray}}
\newcommand{\nn}{\nonumber}

\def\ft#1#2{{\textstyle{\frac{\scriptstyle #1}{\scriptstyle #2} } }}
\def\fft#1#2{{\frac{#1}{#2}}}

\def\0{{\sst{(0)}}}
\def\1{{\sst{(1)}}}
\def\2{{\sst{(2)}}}
\def\3{{\sst{(3)}}}
\def\4{{\sst{(4)}}}
\def\5{{\sst{(5)}}}
\def\6{{\sst{(6)}}}
\def\7{{\sst{(7)}}}
\def\8{{\sst{(8)}}}
\def\sst#1{{\scriptscriptstyle #1}}

\begin{document}
	
	\begin{center}
		{\Large {\bf Gravitational Collapse: Generalizing Oppenheimer–Snyder and a Conjecture on Horizon Formation Time}} 
		
		\vspace{20pt}
		
H. Khodabakhshi\hoch{1}, H. L\"u\hoch{1,2}, F. Shojai\hoch{3}

\vspace{20pt}
		
		{\it \hoch{1}Center for Joint Quantum Studies and Department of Physics,\\
			School of Science, Tianjin University, Tianjin 300350, China }
			\medskip
		
		{\it \hoch{2}The International Joint Institute of Tianjin University, Fuzhou,\\ Tianjin University, Tianjin 300350, China}

\medskip
		{\it \hoch{3}Department of Physics, University of Tehran, P.O. Box 14395-547, Tehran, Iran}

		\vspace{40pt}
		
\underline{ABSTRACT}
	\end{center}

We generalize the Oppenheimer–Snyder model of gravitational collapse by considering a broader class of static, spherically symmetric exterior spacetimes, with an interior geometry described by a Friedmann–Lemaître–Robertson–Walker (FLRW) geometry. Using Painlev\'e--Gullstrand (PG) coordinates for the spatially flat interior geometry ($k=0$) and a Novikov-like coordinate system for the spatially closed geometry ($k=1$), we ensure a smooth transition between the interior and exterior of the collapsing star. By providing general formulas, we analyze how apparent and event horizons form during the collapse and examine whether the matter satisfies standard energy conditions. For both $k=0$ and $k=1$ cases, we study explicit examples such as Schwarzschild, Schwarzschild--AdS/dS, and Reissner--Nordstr\"om (RN) black holes, taking into account the effects of the cosmological constant and electric charge. These factors significantly influence the collapse process and can impose constraints on the physical parameters. Our analysis leads to two important results: First, to form a black hole, there is a minimum or critical initial radius for the star to begin collapsing. Second, we propose a conjecture of an inequality regarding the event horizon formation time, starting from the critical radius,  namely  $\Delta T_{\text{eh}} \le 19M/6$. The upper bound is saturated by the Schwarzschild black hole.

\vfill {\footnotesize  h\_khodabakhshi@tju.edu.cn \ \ \  mrhonglu@gmail.com\ \ \ fshojai@ut.ac.ir}

\thispagestyle{empty}
\pagebreak

\tableofcontents
\addtocontents{toc}{\protect\setcounter{tocdepth}{2}}

\section{Introduction}

General relativity (GR) has been a successful theory for explaining gravity and cosmology, both in theory and through observation. It provides a consistent explanation for many phenomena in the universe and its evolution. One interesting result of GR is the prediction of the existence of black holes. According to the no-hair theorem \cite{1,1*}, a stationary black hole can be fully described by just three properties: its mass, angular momentum, and electric charge. Black holes have unique features such as singularities—points at which the known laws of physics no longer apply \cite{2,2*,2**}. However,  the weak cosmic censorship conjecture (WCCC) \cite{3} states that in a geodesically complete spacetime, matter satisfying the null energy condition cannot form a naked singularity accessible  to an asymptotic observer. A general, model-independent formula for testing the WCCC, based on the gedanken experiments for the Kerr--Newman black hole \cite{Wald:1974hkz,Sorce:2017dst}, was recently proposed in \cite{4,Lu:2025efo}. The strong cosmic censorship conjecture (SCCC) \cite{5}, on the other hand, states that if we know the initial conditions of a system, the laws of physics should allow us to predict its future everywhere in spacetime. This suggests that Cauchy horizons inside black holes are unstable. The existence of black holes has been supported by observations, such as the first image of a black hole’s shadow taken by the Event Horizon Telescope \cite{6,6*,6**}, and the detection of binary black hole mergers by the LIGO and Virgo collaborations \cite{7}. 

In this paper, we study black hole formation in the gravitational collapse of a star by generalizing the classic Oppenheimer–Snyder (OS) model \cite{8,8*} within the framework of standard Einstein gravity. Here, “generalizing” refers to extending the OS matching procedure to a broader class of static, spherically symmetric solutions of Einstein’s equations—such as Reissner–Nordström (RN) or Schwarzschild-de Sitter (dS) and Schwarzschild-Anti de Sitter (AdS)—while keeping the interior as a standard Friedmann--Lemaître--Robertson--Walker (FLRW). The exterior and interior geometries are always assumed to satisfy the Einstein equations with a physically interpretable source (e.g., electromagnetic field or cosmological constant), and and we verify that the interior matter satisfies standard energy conditions to ensure physical validity. Some interesting papers that investigate OS collapse in various cases can be found in the following references \cite{9,9*,10, 10*, 11, r1, r2, r3, r4}. In \cite{r2} the auther studied the OS collapse for the Schwarzschild-AdS/dS black hole and in \cite{r3,r4} the focus is on regular black holes. Earlier work on gravitational collapse with electric charge was carried out by Raychaudhuri~\cite{R2}, who considered self-gravitating charged dust with dynamical interior fields. Our model follows the OS framework, where the star is neutral and collapses into a fixed exterior geometry. We consider a scenario in which the interior of the star is modeled as a homogeneous and isotropic perfect fluid, similar to a FLRW universe, while the exterior spacetime is described by a more general static, spherically symmetric solutions of Einstein's equations—such as Schwarzschild, RN, or Schwarzschild-AdS/dS. Our systematic mathematical analysis of OS collapse enables us to study different physically motivated black hole geometries, where the effective stress-energy tensor associated with the exterior metric is simply the one required by Einstein’s equations. For Schwarzschild-AdS/dS spacetime, the effective exterior source is a perfect fluid with isotropic pressure $p = -\Lambda/8\pi$; for the RN case, it is anisotropic. Both arise naturally in GR: the former from the cosmological constant, the latter from the electromagnetic field.
We shoudl also mention that the WCCC holds during the collapse. It has been shown that the WCCC holds true for all spherically symmetric and static solutions \cite{4}. We consider both spatially flat ($k=0$) and closed ($k=1$) FLRW models for the interior geometry and match them smoothly to the exterior spacetime, according to the Israel junction conditions \cite{Israel:1966rt}.

In Section 2, to facilitate a smooth joining of the two geometries at the surface of the star, we use Painlev\'e--Gullstrand (PG) coordinates \cite{12,13,14} for the $k=0$ case and a Novikov-like coordinate system inspired by Novikov’s approach \cite{15} for the $k=1$ case. These coordinate systems are adapted to freely falling observers and help us avoid coordinate singularities on the horizon. The use of PG coordinates for OS and more general dust collapse has been explored in previous works: Ref.~ \cite{K2} analyzed the OS collapse of a $k=0$ FLRW interior in PG coordinates, while Ref.~ \cite{K3} extended this framework to the broader class of Lema\^itre--Tolman--Bondi (LTB) solutions, which include both $k=0$ and $k=1$ FLRW as special cases. By expressing both the interior and exterior metrics in PG and Novikov-like coordinate systems, we directly compare them at the stellar surface boundary. This enables us to derive the equations governing the motion of the stellar surface during collapse and to obtain  physical quantities such as the density and pressure.  We also derive the junction conditions for a smooth connection between two spacetimes. These conditions require not only the continuity of the metric across the boundary but also the continuity of the extrinsic curvature \cite{2}. This confirms that the two geometries can be joined smoothly without introducing any thin-shell matter to the surface, thus ensuring a physically consistent description of the collapsing star. It should be noted that the study of OS collapse is independent of the chosen coordinate system. All information about collapse, including horizon dynamics and stellar matter properties, can be derived using any coordinate system. Here, however, we use a unified coordinate system for both inside and outside the star. This simplifies the interpretation of the results and enables generalization to arbitrary black hole geometries.

In Section 3, we study the dynamics of gravitational collapse with a focus on the formation and evolution of apparent and event horizons during the collapse of a star. We employ the standard terminology for ingoing and outgoing null geodesics to derive the relevant formulas for apparent and event horizons \cite{n1,n2,n3}. We find that the apparent horizon emerges within the star, defining the boundary of the region containing the trapped surfaces. To ensure the physical validity of the matter content during collapse, we analyze the energy conditions — namely, the weak, null, dominant, and strong energy conditions. For $k = 0$ (spatially flat interior geometry) and $k = 1$ (spatially closed interior geometry), we derive the general form of the equations describing the collapse under the assumption of static and spherically symmetric exterior solutions. These results enable us to consider different examples and compare the influence of spacetime geometry on the formation and behavior of horizons during black hole formation.

In Section 4, we examine explicit examples of black hole formation during the gravitational collapse of a star, considering both the $k = 0$ and $k = 1$ cases. Since real stars are complex—they contain pressure and other forces that influence their collapse—to make our model more physically relevant, we include the effects of a cosmological constant and electric charge. We begin by studying Schwarzschild-AdS/dS black holes, in which the presence of the cosmological constant $\Lambda$ alters the behavior of the horizons \cite{r2}. For $k = 0$, a consistent OS collapse exists for both the Schwarzschild-AdS and Schwarzschild-dS cases, but with a restriction on the allowed range of $|\Lambda|$. We determine this range for given the mass of the black hole. However, we show that no consistent OS collapse exists for Schwarzschild-AdS/dS black holes in the $k = 1$ case when the star begins collapsing from rest (i.e., with zero initial velocity). In this case, a consistent OS collapse can still be achieved for Schwarzschild-dS spacetime if the star starts collapsing with a nonzero initial velocity. We also examine the collapse of a star into a RN black hole \cite{R2}. While related studies of charged collapse was considered in \cite{R2}, our analysis focuses on the evolution of horizons, energy conditions, and the minimum initial radius required for black hole formation within the OS framework—features that have not been addressed in this context for both $k = 0$ and $k = 1$ cases. We find that only when the black hole is sufficiently close to the extremal limit, the inner apparent horizon connects the inner and outer horizons inside the star. Furthermore, due to the electric charge, gravity becomes (infinitely) repulsive near the center. As a result, the star does not collapse all the way to zero size ($R = 0$), but instead reaches a finite minimum radius. In the appendix, to verify our general formulas for the simplest example, we revisit the Schwarzschild black hole and compute how the stellar radius evolves in time, as well as how the apparent and event horizons form and evolve during collapse. These results are used in Section 5, where we present a conjecture on the horizon formation time. In all examples, we also check whether the matter inside the star satisfies the standard energy conditions. This ensures that our model represents physically valid scenarios within general relativity.

In Section 5, we investigate the time required for the formation of the event horizon during gravitational collapse, denoted as $ \Delta T_{\text{eh}} $. We address two key questions: (1) Is there a minimum initial radius $ R_0 $ from which the collapse must begin in order to form a black hole? and (2) Is $ \Delta T_{\text{eh}} $ a constant that depends only on the black hole parameters? We analyze these questions for both spatially flat ($ k = 0 $) and closed ($ k = 1 $) geometries. We first show that a minimum critical initial radius, $ R_{0,\text{min}} $, exists. Furthermore, we find that $\Delta T_{\rm eh}$ is the same as the minimum critical proper time measured via the collapsing stellar surface. We propose a conjecture that the collapsing of the Schwarzschild black hole sets an upper bound of  $\Delta T_{\rm eh}$ for the OS collapse. 

In this paper, we employ PG coordinates and derive general formulas for the evolution of the stellar radius, as well as for the event and apparent horizons, and we systematically analyze the energy conditions in the context of OS collapse—results that have not been presented together in previous studies \cite{r2, R2}. We consider both the $k=0$ case (flat geometry) and the $k=1$ case (closed geometry), and for $k=1$ we introduce a new PG-like coordinate system. This comprehensive framework enables us to formulate a conjecture on the horizon formation time, which has not been previously discussed in the literature.

Finally, in Section 6, we summarize the key findings and conclude the paper.

\section{Gravitational collapse of a star}
 
We aim to describe the gravitational collapse of a star according to the OS model~\cite{8,8*} within the framework of standard Einstein gravity. In the standard OS scenario, the stellar matter is modeled as a homogeneous and isotropic perfect fluid. Depending on the initial conditions of the star, the interior geometry is governed by a spatially flat FLRW metric ($k = 0$), while the exterior geometry is described by the Schwarzschild black hole metric as the vacuum solution. 

In this paper, we extend this setup in two directions:  One is to consider also the closed FLRW metric $(k=1$) for the interior geometry. The other is to replace the exterior Schwarzschild geometry with a more general static, spherically symmetric spacetime given by the metric ansatz
\begin{equation}\label{met}
	ds^2 = -f(r) dt^2 + \frac{dr^2}{f(r)} + r^2 d\Omega^2.
\end{equation}
Substituting (\ref{met}) into the Einstein equations $ G_{\mu}^\nu = 8\pi T_\mu^\nu $ (with $ G = 1 $), the energy density and anisotropic pressures of the exterior matter content are derived as
\begin{align}\label{eom}
	\rho &= -p_r = -\frac{1}{8\pi} \frac{-1 + f(r) + r f'(r)}{r^2}, \\
	p_{\theta} &= p_{\phi} = \frac{f'(r)}{r} + \frac{f''(r)}{2}.
\end{align}
The function $f(r)$ corresponds to a known, physically motivated solution of Einstein’s equations. Furthermore, the interior matter satisfies standard energy conditions, ensuring the collapse describes a physically valid scenario within general relativity. We consider a compact star embedded in the regular spacetime defined by (\ref{met}). The star does not necessarily reside in vacuum; rather, it represents a region that is significantly denser than its (static) surroundings and thus undergoes gravitational collapse. The stress-energy tensor of the stellar matter occupying this dense region is generally unknown. However, it can be determined directly by demanding a smooth transition of the metric across the boundary of the star. To achieve this, we adopt the FLRW metric for the interior geometry and choose the time coordinate of the exterior (interior) metric to correspond to the proper time of freely falling (comoving) observers. In this section we will proceed by smoothly matching the two geometries for both spatially flat ($k = 0$) and closed ($k = 1$) FLRW universes. We will demonstrate how the core geometry is smoothly joined to the exterior spacetime \cite{2} without requiring the introduction of a thin-shell distribution of matter at the surface.
 
\subsection{$k=0$ case}

Here, in the case where $ k = 0 $, we consider the gravitational collapse of a star according to the OS model \cite{8,8*}. To do so, it is convenient to write the regular metric (\ref{met}) in terms of PG coordinates \cite{12,13,14}, which are adapted to the freely falling radial observer starting from rest at infinity. Performing the following coordinate transformation on $t$ 
\be\label{ttran}
dt \rightarrow d\tau-\frac{\sqrt{1-f}}{f} dr,
\ee
we can eliminate the horizon coordinate singularity and rewrite the metric (\ref{met}) as
\be\label{metpg}
ds^2 =-d\tau^2 + (dr+\sqrt{1-f} d\tau)^2 + r^2 d\Omega^2.
\ee
The interior geometry of the collapsing star can be described by the spatially flat FLRW cosmological model, {\it i.e.}
\be\label{frw}
ds^2 =-d\tau^2 + a(\tau)^2 (dr_c^2 + r_c^2 d\Omega^2).
\ee
which can also be written into the PG coordinates, namely
\be\label{frwpg}
ds^2 = -d\tau^2 + (dr - r H(\tau) d\tau)^2 + r^2 d\Omega^2,
\ee
where $r(\tau,r_c) = a(\tau)r_c$, $H(\tau,r_c) = \dot{r}(\tau,r_c)/r(\tau,r_c)$, $a(\tau)$ is the scale factor and $r_c$ is the comoving radial coordinate.

For the radial geodesic of a test particle, we have $d\Omega^2=0$ and $ds^2=-d\tau^2$. From eqs. (\ref{metpg}) and (\ref{frwpg}) on the star surface, $R(\tau)=r(\tau,r_{c,s}=r_{\text{surface}})$, we obtain
 \be\label{req}
\frac{dR}{d\tau}=-\sqrt{1-f(R)}=R H=\dot{R} \hspace{1cm} \rightarrow \hspace{1cm} \frac{\dot{R}^2}{R^2}=\frac{1-f(R)}{R^2}. 
 \ee
It should be noted that the choice of the geodesic $ dR + \sqrt{1 - f} \, d\tau = 0 $ corresponds to a freely falling observer starting from rest at infinity. This ensures $ \tau $ is the proper time of such a global infalling frame, providing a regular, horizon-crossing foliation that synchronizes the interior and exterior regions of the star. While other geodesics exist, this choice simplifies the matching and avoids coordinate singularities.
Substituting $R(\tau) = a(\tau) r_{\text{surface}}$ into the Friedmann equation (for $G=1$) gives
 \be\label{freq}
\frac{\dot{R}^2}{R^2}=\frac{8\pi}{3} \rho(R).
\ee
Now comparing (\ref{req}) with (\ref{freq}), one finds that the metric is the same on both sides of the star if the density inside the star is
\be\label{den}
\rho(\tau)=\frac{3}{4\pi} \frac{1-f(R(\tau))}{2R(\tau)^2}.
\ee
Taking the time derivative of (\ref{den}) and substituting it into the continuity equation of the total perfect fluid filling the interior of the star, $\dot{\rho} + 3 \dot{a}/ a ( \rho + p ) = 0$, gives the stellar surface pressure as
\be\label{pre}
p(\tau)=\frac{1}{8\pi} \frac{-1+f(R(\tau))+R(\tau) f'(R(\tau))}{R(\tau)^2}.
\ee
where the prime denotes the derivative with respect to $R$. The density (\ref{den}) and the isotropic pressure (\ref{pre}) represent the total energy density and pressure on the surface of the star, which consist of the background part given by (\ref{eom}) and an unknown stellar part. Additionally, equations (\ref{den}) and (\ref{pre}) describe the parametric equation of state for the stellar matter, since knowing $R(\tau)$, both the pressure and density can be expressed as functions of the proper time of a particle located on the surface of the star.

As mentioned earlier, we assume that any test particle on the stellar surface is a freely falling particle, and hence its radial trajectory obeys (\ref{req}). To clarify the motion of the boundary, we note that using eqs.~(\ref{eom}) and (\ref{pre}), which describe the radial pressure outside and inside the star respectively, one finds that the net radial pressure jump at the surface vanishes, i.e.,
	$
	p - p_r \big|_{r=R} = 0.
	$
This condition indicates that there is no discontinuity in radial pressure across the boundary. In our setup, combined with the junction conditions for the metric and extrinsic curvature, this implies that the boundary particles experience no four-acceleration. As a result, their worldlines follow radial timelike geodesics, consistent with eq.~(\ref{req}).

So far we have employed a single coordinate system (PG) for both the interior and exterior regions to ensure the continuity of the metric. On the stellar surface, these metrics are matched, and this process directly yields the evolution equation of the stellar surface (\ref{req}). This approach simplifies the analysis by avoiding complications arising from coordinate transformations at the boundary. However, it is important to verify the junction conditions for matching the interior and exterior geometries of the star; they are satisfied through two key requirements: the continuity of the metric and the continuity of the extrinsic curvature. The latter ensures a smooth connection between the interior and exterior geometries. Therefore, we will verify the smooth joining of the two geometries by calculating the components of the extrinsic curvature for the interior and exterior metrics. Using eqs. (\ref{met}) and (\ref{frw}), a straightforward calculation \cite{2} yields the components of the extrinsic curvature on the stellar surface, $r(\tau) = R(\tau)$, as
\bea\label{kcom} 
&{ ^{(\text{in})}K^\tau_\tau} = 0, \quad \quad { ^{(\text{in})}K^\theta_\theta} ={ ^{(\text{in})}K^\phi_\phi} = \frac{1}{R}, \nonumber \\
&{ ^{(\text{out})} K^\tau_\tau} = \frac{\partial_\tau \sqrt{\dot{R}^2 + f(R)}}{\dot{R}}, \quad \quad{ ^{(\text{out})} K^\theta_\theta} = { ^{(\text{out})} K^\phi_\phi} = \frac{\sqrt{\dot{R}^2 + f(R)}}{R}.
\eea
This requires $\sqrt{\dot{R}^2+f(R)}=1$, which yields the star surface equation (\ref{req}). Moreover, from equation (\ref{den}), one can calculate the total mass $m$ of the star given the form of the function $f(R)$.  By comparing the interior Friedmann equation with the geodesic equation of radial particles in the exterior spacetime, we derive the mass relation. Notably, this mass relation—or equivalently, the geodesic equation—can be interpreted as the condition enforcing the continuity of the extrinsic curvature across the stellar surface.

\subsection{$k=1$ case}

In the case of a closed universe, there is a maximum value for the scale factor $a=a_{\text{max}}$, corresponding to a maximum radius $R=R_{\text{max}}$. At this radius, the collapse begins with zero initial velocity $\dot{R}|_{R=R_{\text{max}}}=\dot{a}|_{a=a_{\text{max}}}=0$. To describe this scenario, we express the metric (\ref{met}) in coordinates that generalize the Novikov-like case \cite{15} \footnote{This form of the metric differs from that presented in Novikov's paper \cite{15}. Since our formulation was inspired by Novikov's approach, we refer to the coordinate system as Novikov-like.}, which is compatible with freely falling radial particles starting from rest at a finite radius $r_0$, instead of using PG coordinates \cite{12,13,14}. Using the following coordinate transformation on $t$ 
\be\label{ttrann}
dt \rightarrow d\tau-\frac{\sqrt{1-f(r)/f(r_0)}}{f(r)/f(r_0)} dr,
\ee
the metric takes the form
\be\label{npg}
ds^2 =-d\tau^2 + \frac{(dr+\sqrt{f(r_0)-f(r)} d\tau)^2}{f(r_0)} + r^2 d\Omega^2.
\ee
and $r_0$ labels the initial radial position of the particles. In the case of an asymptotically flat metric, $f(r_0)\rightarrow 1$  in the limit of $r_0\rightarrow \infty$, and one obtains the Schwarzschild metric in PG coordinate (\ref{metpg}).

In Novikov-like coordinates, the FLRW metric for the case $k=1$ is expressed as 
\be\label{frwnpg}
ds^2 = -d\tau^2 + \frac{(dr - r H(\tau) d\tau)^2}{1-r_c^2} + r^2 d\Omega^2.
\ee

Following the same procedure, the radial geodesic of a test particle is given by
 \be\label{reqc}
\frac{dR}{d\tau}=-\sqrt{f(R_0)-f(R)}=R H=\dot{R} \hspace{1cm} \rightarrow \hspace{1cm} \frac{\dot{R}^2}{R^2}=\frac{f(R_0)-f(R)}{R^2}. 
\ee
Here, $R_0$ denotes the initial radius of the collapsing star at $\tau = 0$, where $R_0 = r_0$ at the boundary to ensure a consistent physical setup for the collapse.
Substituting $R(\tau)=r(\tau,r_{c,s}=r_{\text{surface}})$ into the Friedmann equation (for $G=1$ and $k=1$) yields
\be\label{cfreq}
\frac{\dot{R}^2}{R^2}=\frac{8\pi}{3} \rho(R)-\frac{r_{c,s}^2}{R^2}.
\ee
Comparing (\ref{reqc}) with (\ref{cfreq}), we find that the metrics (\ref{npg}) and (\ref{frwnpg}) are identical on both sides of the star if the density inside the star is given by (\ref{den}), and the surface comoving radial coordinate $r_c$ is defined as
\be\label{rc}
r_{c,s}^2=1-f(R_0).
\ee

Here, we do not consider $k = -1$ because it leads to inconsistencies in the equations of motion. To see this, consider eqs.~(16) and (17). Equation~(17) is derived for $k=1$; if we instead consider $k=-1$, the coefficient of the $r_c^2/R^2$ term on the right-hand side becomes positive, and from eq.~(16) we obtain
\begin{align}\label{mk}
	\dot{R}^2 = f(R_0) - f(R) \quad \Rightarrow \quad \dot{R}^2 + \frac{2}{R_0} = \frac{2}{R}.
\end{align}
Meanwhile, eq.~(17) with $k=-1$ gives
\begin{align}\label{mk1}
	\dot{R}^2 - r_{c,s}^2 = \frac{8\pi}{3} \rho(R) R^2.
\end{align}
eqs.~\eqref{mk} and \eqref{mk1} are inconsistent in the $k=-1$ case. In contrast, for $k=0$ and $k=1$, the equations are consistent in the limits $R_0 \to \infty$ and finite $R_0$, respectively.

It is important to note that the functional forms of the pressure and density remain the same as in the $k=0$ case. Consequently, the total mass of the star remains unchanged.  Furthermore, by considering $\dot{R}=0$, where $R=R_{\text{max}}$, implies $f(R_0)=f(R_{\text{max}})$ or equivalently $R_0=R_{\text{max}}$. Using (\ref{rc}) and the Friedmann equation (\ref{cfreq}), this can be expressed in terms of the density as
\be\label{denm}
\rho(R_{\text{max}})=\frac{3}{4\pi} \frac{1-f(R_{\text{max}})}{2R_{\text{max}}^2}.
\ee
Same as $k=0$ case, we can similarly verify the smooth joining of the two geometries by calculating the components of the extrinsic curvature for the interior and exterior metrics. Since we have changed the interior geometry to the  $k=1$ case, the components of the extrinsic curvature derived from the metric (\ref{frwnpg}) can be calculated as
\bea\label{kcomn}
&^{(\text{in})} K^\tau_\tau = 0, \quad \quad ^{(\text{in})} K^\theta_\theta = ^{(\text{in})} K^\phi_\phi = \frac{\sqrt{1-r_{c,s}^2}}{R}.
\eea
Comparing these with the components of the exterior extrinsic curvature, we obtain the requirement
\be
\sqrt{\dot{R}^2+f(R)}=\sqrt{1-r_{c,s}^2}.
\ee
Substituting eq. (\ref{rc}) into this condition directly leads to the surface evolution equation (\ref{reqc}).

Later we will consider explicit examples using the generalized OS gravitational collapse of a star to the Schwarzschild-AdS/dS black hole for both $k=0$ and $k=1$ cases.

\section{The dynamics of black hole creation}

During the gravitational collapse of a star, energy conditions and the formation of horizons play a crucial role in determining the physical validity of the collapsing system and understanding the dynamics of black hole creation. In the following sections, we will discuss these topics in detail.

\subsection{Energy conditions}

For an an isotropic perfect fluid, the energy conditions are constraints on the energy density $\rho$ and pressure $p$. These conditions ensure that stellar matter behaves in a physically reasonable way during collapse. In the exterior, the solution is static with the energy density and pressures given by \eqref{eom}. In the interior, the radial pressure near the surface is the same as its exterior neighbour, by the virtual of the OS formalism; however, the energy density is different. It is thus necessary to check the energy conditions of the interior collapsing matter. The positivity of the energy density ($\rho \geq 0$) in eq. (\ref{den}) requires $f(R) \leq  1$, which is true for all asymptotically flat spacetimes. Assuming a positive energy density, both the Weak Energy Condition (WEC) and the Null Energy Condition (NEC) require $p + \rho \geq 0$, with equality applying only to the NEC. To satisfy the Dominant Energy Condition (DEC), which requires $|p| \leqslant   \rho$, an additional condition $p - \rho \leq 0$ must also be met. The DEC includes both the WEC and the NEC. Furthermore, the Strong Energy Condition (SEC) is defined as $p + \rho \geq 0$ and $\rho + 3p \geq 0$. For the given density and pressure in eqs. (\ref{den}) and (\ref{pre}), one can easily check the energy conditions. The definitions of $\rho$ and $p$ remain the same for both the $k = 0$ and $k = 1$ cases. In section 4, we will investigate these conditions for Schwarzschild-AdS/dS black holes.

\subsection{Horizons}

We now examine the evolution of event and apparent horizons during gravitational collapse. We will show that both the event horizon and the apparent horizon are initially absent. As the collapse progresses, these boundaries form due to the increasing strength of gravity. The event horizon starts from zero size and grows until it reaches the black hole event horizon (e.g., $R_+ = 2M$ in the Schwarzschild case). Meanwhile, the apparent horizon forms while the star's radius is still shrinking and eventually aligns with the star's surface as it collapses to zero size. This means that if the black hole outside the star is singular, the star's radius and the apparent horizon radius will be zero at the end of the collapse. However, if the black hole is regular, such as Hayward or Bardeen, neither the surface nor the apparent horizon will reach zero \cite{9}. In what follows, we will derive a general formula for the star's radius and study the evolution of the event and apparent horizons for both the $k = 0$ and $k = 1$ cases.

\subsubsection{$k = 0$ Case}

In this case, we assume the star begins collapsing at $\tau = 0$ from an arbitrary initial radius $R = R_0$ with a non-zero initial velocity. The initial zero velocity is only possible at an infinite radius from the star, however, starting the collapse from such a radius would be unphysical. Therefore, the collapse must begin from a finite radius with a non-zero initial velocity. Moreover, the collapse velocity never actually reaches zero at any point during the collapse process. Otherwise, according to the Israel matching conditions (\ref{kcom}), the interior FRW cannot correspond to the flat $k=0$ case. The evolution of the star's surface radius $R$ is governed by eq. (\ref{req}), where the Hubble parameter $H$ is negative. This gives
\begin{equation}\label{tsr}
	T(R) = \int_0^T d\tau = \int_{R_0}^R -\frac{dR}{\sqrt{1 - f(R)}}.
\end{equation}
Inside the star, radial null geodesics are given by $dr/d\tau = -1 + rH$ and $dr/d\tau = 1 + rH$, as derived from eq. (\ref{frw}). Since $H < 0$, the first equation corresponds to ingoing null geodesics \cite{n1,n2,n3}, while the second represents both ingoing and outgoing null geodesics depending on whether $r > -1/H$ or $r < -1/H$, respectively. This implies the presence of trapped surfaces with radii smaller than $-1/H$, and so the interior apparent horizon is given by
\begin{equation}\label{rap}
	R_{\text{ap}}(R) = \frac{R}{\sqrt{1 - f(R)}},
\end{equation}
where $H(R) = -\sqrt{1 - f(R)}/R$ from eq. (\ref{req}). It is clear that $R_{ap}(R)$ joins $R$ at $R_+$ since $f(R_+)=0$. Solving this equation yields $R$ as a function of $R_{\text{ap}}$, which can then be substituted into eq. (\ref{tsr}) to obtain $T_{\text{ap}}(R)$. The equation describing the apparent horizon surface is:
\begin{equation}\label{seq}
	F(r, \tau) = r - \frac{R(\tau)}{\sqrt{1 - f(R(\tau))}},
\end{equation}
with the normal vector of the apparent horizon given by $n_\alpha = -\partial_\alpha F = (-x, -1, 0, 0)$. Using eqs. (\ref{metpg}) and (\ref{req}), $n^2$ is obtained as
\begin{equation}\label{n2}
	n^2 = -x^2 + 2x, \quad \quad \quad x \equiv 1 - \frac{R f'(R)}{2(1 - f(R))}.
\end{equation}
On the hypersurface $RH = -1$, $n^2 > 0$ indicates a timelike apparent horizon, while $n^2 < 0$ indicates a spacelike apparent horizon. Usually in the case of dynamical horizons—such as in the OS collapse of stars into non-singular black holes [22] and into Einstein-Gauss-Bonnet black holes [24]—the apparent horizon inside the star is actually a timelike hypersurface.

To determine the interior event horizon $R_{\text{eh}}$, we consider outgoing null geodesics \cite{n1,n2,n3}that terminate on the star's surface at the future event horizon of the outer spacetime. Since the exterior geometry is known (e.g., Schwarzschild or RN), the outer event horizon $R_+$ is determined globally by the asymptotic structure. We use the condition $R_{\text{eh}}(R_+) = R_+$ as a boundary condition, meaning that the interior null geodesic matches continuously onto the exterior event horizon at $R = R_+$. These outgoing null geodesics in the interior satisfy $\dot{R}_{\text{eh}} = 1 + R_{\text{eh}} H$, which can be rewritten in terms of $R$ as
\begin{equation}\label{reh}
	\frac{dR_{\text{eh}}(R)}{dR} = \frac{R_{\text{eh}}(R)}{R} - \frac{1}{\sqrt{1 - f(R)}},
\end{equation}
where the derivation follows from $ds^2 = 0$ for null paths in the interior metric~\eqref{frwnpg}, with $d\Omega = 0$. Solving this equation inward from $R = R_+$ gives $R_{\text{eh}}(R)$, the location of the interior section of the event horizon. Substituting $R$ as a function of $R_{\text{eh}}$ into Eq.~\eqref{tsr} then yields $T_{\text{eh}}(R)$.

\subsubsection{$k = 1$ Case}

For $k = 1$, where the star begins collapsing at $\tau = 0$, we consider two distinct scenarios: (i) $R_0 = R_{\text{max}}$, where collapse begins from rest (i.e., with zero initial velocity), and (ii) $R_0 < R_{\text{max}}$, where the initial velocity is non-zero. The evolution of the star's surface radius $R$ is governed by eq. (\ref{reqc}), yielding
\begin{equation}\label{tsrc}
	T(R) = \int_0^T d\tau = \int_{R_{\text{max}}}^R -\frac{dR}{\sqrt{f(R_{\rm max}) - f(R)}}.
\end{equation}
In this case, radial null geodesics \cite{n1,n2,n3} are given by $dr/d\tau = -\sqrt{f(R_{\text{max}})} + rH$ and $dr/d\tau =\sqrt{ f(R_{\text{max}})} + rH$, as derived from eq. (\ref{frwnpg}). The interior apparent horizon is given by
\begin{equation}\label{rapc}
	R_{\text{ap}}(R) = \frac{\sqrt{f(R_{\text{max}})} R}{\sqrt{f(R_{\text{max}}) - f(R)}},
\end{equation}
where $H(R) = -\sqrt{f(R_{\text{max}}) - f(R)}/R$. Solving this equation provides $R$ as a function of $R_{\text{ap}}$, which can then be substituted into eq. (\ref{tsr}) to obtain $T_{\text{ap}}(R)$. The apparent horizon surface is described by
\begin{equation}\label{seqc}
	F(r, \tau) = r - \frac{\sqrt{f(R_{\text{max}})}R(\tau)}{\sqrt{f(R_{\text{max}}) - f(R(\tau))}},
\end{equation}
with the normal vector of the apparent horizon given by $n_\alpha = -\partial_\alpha F = (-y, -1, 0, 0)$. The quantity $n^2$ is expressed as
\begin{equation}\label{n2c}
	n^2 =1- (y-\sqrt{f(R_{\text{max}})})^2 f(R_{\text{max}}), \quad  \quad y \equiv 1 - \frac{Rf'(R)}{2(f(R_{\text{max}}) - f(R))}.
\end{equation}
On the hypersurface $RH = -\sqrt{f(R_{\text{max}})}$, $n^2 > 0$ indicates a timelike apparent horizon, while $n^2 < 0$ indicates a spacelike apparent horizon.

To determine the interior event horizon $R_{\text{eh}}$, we again consider outgoing null geodesics \cite{n1,n2,n3} that reach the black hole horizon, satisfying $R_{\text{eh}}(R_+) = R_+$. These geodesics obey $\dot{R}_{\text{eh}} = \sqrt{f(R_{\text{max}})} + R_{\text{eh}}H$, which can be rewritten as
\begin{equation}\label{rehc}
	\frac{dR_{\text{eh}}(R)}{dR} = \frac{R_{\text{eh}}(R)}{R} - \sqrt{\frac{f(R_{\text{max}})}{f(R_{\text{max}}) - f(R)}}.
\end{equation}
Solving this equation gives $R_{\text{eh}}(R)$, and substituting $R$ as a function of $R_{\text{eh}}$ into eq. (\ref{tsr}) yields $T_{\text{eh}}(R)$. In the next section, we will investigate the horizons evolution for the Schwarzschild-AdS/dS black holes.

In the case $k = 1$, the star can also begin collapsing at $\tau = 0$ with an initial velocity and initial radius $R = R_0$, where the evolution of the star's surface radius $R$ is given by Eq.~(\ref{reqc}) as  
\begin{equation}\label{tsrc0}
	T(R) = \int_0^T d\tau = \int_{R_0}^R -\frac{dR}{\sqrt{f(R_{\rm max}) - f(R)}}.
\end{equation}
It should be noted that the evolution of the apparent and event horizons is still given by eqs.~(\ref{rapc}) and (\ref{rehc}), respectively.

\section{Explicit examples}

We now examine physically interesting examples to verify the general formalism developed in the previous sections. In Appendix A, we present the analysis for the simplest case—the Schwarzschild black hole—where its results are used in Section 5 to discuss horizon formation time. In this section, we first consider OS collapse into Schwarzschild-AdS/dS black holes, followed by an investigation of the RN black hole. Our analysis comprehensively covers both the spatially flat ($k=0$) and closed ($k=1$) interior geometries.

\subsection{Schwarzschild-AdS/dS black holes}

We extend the OS model to Schwarzschild-AdS/dS spacetimes, deriving exact expressions for the collapse dynamics, horizon evolution, and associated physical constraints. For Schwarzschild-AdS/dS black holes, we have
\begin{equation}\label{sch1}
	f(R) = 1 - \frac{2}{R} - \frac{\Lambda}{3}R^2.
\end{equation}
On the horizon $R = r_0$, $f(r_0) = 0$ gives
\begin{equation}\label{lambda}
	\Lambda = \frac{3 (r_0 - 2)}{r_0^3}.
\end{equation}
From this equation, we see that:
- If $r_0 = 2$, then $\Lambda = 0$.
- If $r_0 > 2$, then $\Lambda > 0$, which corresponds to Schwarzschild-dS black holes.
- If $r_0 < 2$, then $\Lambda < 0$, which corresponds to Schwarzschild-AdS black holes.

For $\Lambda > 0$, there are two solutions for $r_0$: $r_0 = R_+$ (event horizon) and $r_0 = R_c$ (cosmological horizon). For Schwarzschild-dS black holes, the star must start collapsing inside the cosmological horizon, where $R_0 < R_c$. We will study these cases in detail below. In \cite{r2}, some aspects of the OS collapse for Schwarzschild-AdS/dS black holes are studied. However, we will investigate new features within the PG coordinate framework, such as the evolution of event and apparent horizons, and we analyze the energy conditions in the context of OS collapse.

\subsubsection{$k = 0$ case}

In the Schwarzschild-dS case, if $r_0 > 2$, there are two horizons: the event horizon ($r_0 = R_+$) and the cosmological horizon ($R_0 < R_c$). In this geometry, the presence of the cosmological constant tends to enlarge the black hole horizon compared to the asymptotically flat case, while the black hole mass tends to reduce the size of the cosmological horizon relative to pure de Sitter space. For the Schwarzschild-AdS case where $r_0 < 2$, there is only one horizon ($r_0 = R_+$).

To find the star's surface radius for Schwarzschild-AdS/dS, we substitute eq. (\ref{sch1}) into eq. (\ref{tsr}), which gives
\bea\label{tsch1}
r_0>2:&&T(R) = \fft{2r_0^{3/2}}{3\sqrt{r_0-2}}\left(\text{arcsinh}\Big(\sqrt{\fft{R_0^3}{r_0^3}(\ft12 r_0-1)}\Big) - \text{arcsinh}\Big(\sqrt{\fft{R^{3}}{r_0^{3}} (\ft12 r_0-1)}\Big)\right),\nn\\
r_0<2:&& T(R)=\fft{2r_0^{3/2}}{3\sqrt{2-r_0}}\left(\text{arcsin}\Big(\sqrt{\fft{R_0^{3}}{r_0^{3}}(1-\ft12 r_0)}\Big) - \text{arcsin}\Big(\sqrt{\fft{R^3}{r_0^3}(1-\ft12 r_0)}\Big)\right).
\eea
Note that we have $r_0=R_+\le R\le R_0$. Unlike in the Schwarzschild case, $R_0$ is bounded above for both $\Lambda > 0$ and $\Lambda < 0$ due to the global spacetime structure. For positive $\Lambda$, $R_0$ should be less than the cosmic horizon radius, {\it i.e.}
\be
R_0 < r_{\rm cos} = \frac{r_0 \left(\sqrt{\left(r_0-2\right) \left(r_0+6\right)}-r_0+2\right)}{2 \left(r_0-2\right)}\,.
\ee
For negative $\Lambda$, the reality condition of the arcsin function requires that
\be\label{r0ads}
R_0 \le \frac{r_0}{(1-\fft12 r_0)^{\fft13}}\,.
\ee
In both cases, $R_0$ should be bigger than a certain $R_{0,\rm min}$ that is to be determined by the condition of the horizon formation.

To calculate the apparent horizon, substituting eq. (\ref{sch1}) into eq. (\ref{rap}), we find
\begin{equation}\label{rapsch1}
	R_{\text{ap}}(R) = \frac{R^{3/2}}{\sqrt{2 + \frac{R^3 (R_+ - 2)}{R_+^3}}},
\end{equation}
where we have replaced $\Lambda$ using eq. (\ref{lambda}) in terms of $r_0 = R_+$. We want to check whether the apparent horizon is timelike or spacelike. To do this, we substitute eq. (\ref{sch1}) into eq. (\ref{n2}), which gives
\begin{equation}\label{n2y}
	y = \frac{3 R_+^3 \left(2 R^3 (-2 + R_+) + R_+^3\right)}{\left(R^3 (-2 + R_+) + 2 R_+^3\right)^2}.
\end{equation}
This expression is always positive for $R_+ > 2$, which means the apparent horizon is timelike during the gravitational collapse for the Schwarzschild-AdS case. For the evolution of the event horizon, substituting eq. (\ref{sch1}) into eq. (\ref{reh}) and using eq. (\ref{lambda}), we get
\begin{align}\label{reh1}
	R_{\text{eh}}(R) = R \bigg( 1 &+ \sqrt{2 R_+} \, \text{Hypergeometric2F1}\bigg[\frac{1}{6}, \frac{1}{2}, \frac{7}{6}, 1 - \frac{R_+}{2}\bigg]\nonumber\\
	&- \sqrt{2 R} \, \text{Hypergeometric2F1}\bigg[\frac{1}{6}, \frac{1}{2}, \frac{7}{6}, -\frac{R^3 (R_+ - 2)}{2 R_+^3}\bigg] \bigg),
\end{align}
where the integration constant is fixed by the condition $R_{\text{eh}}(R_+) = R_+$.

The critical minimum $R_{0,\rm min}$ can be determined by \eqref{reh1}, but a general analytical expression is unlikely. For sufficiently small cosmological constant, a perturbative analytical expression can be achieved. We can define $r_0-2 = \epsilon\ll 1$, so that $\Lambda = 3\epsilon/8 + {\cal O}(\epsilon^2)$. In this case, we have
\bea
T(R) &=& \frac{\sqrt{2}}{3} \left( (R_0^{3/2}-R^{3/2}) + \frac{1}{96}\left( R^{9/2} - R_0^{9/2} \right)\epsilon\right) + {\cal O}(\epsilon^2),\label{tsch1e}\\
R_{\text{ap}}(R) &=&  \frac{R^{3/2}}{\sqrt{2}} - \frac{R^{9/2} }{32 \sqrt{2}}\epsilon + {\cal O} (\epsilon^2) \,,\label{raph1e}\\
R_{\text{eh}}(R) &=&  \left(3R - \sqrt{2} \, R^{3/2} \right) + \frac{1}{224} \left(R \left(96 + \sqrt{2} \, R^{7/2} \right) \right) \epsilon + {\cal O}(\epsilon^2)\label{reh1e} .
\eea
Thus we find that
\be
R_{0,\rm min}=\fft92 + \frac{8865}{1792} \epsilon + {\cal O}(\epsilon^2),\qquad
\Delta T_{\rm eh} =\frac{19}{6}+\frac{36451}{16128} \epsilon + {\cal O}(\epsilon^2)\,,\label{deltaTlinearep}
\ee
where $\Delta T_{\text{eh}}$ denotes the black hole formation time, given by Eq.~\eqref{ttt}, and $R_{0,\text{min}}$ is the minimum initial radius derived from Eq.~\eqref{r0}.

When $|\Lambda|$ is large, the requirement of the existence of the critical $R_{0,\rm min}$, and furthermore it must be less than $r_{\rm cos}$ for positive $\Lambda$ and $T(R_{0,\rm min})$ must be real put a severe constraint on $\Lambda$. (Recall that we have fixed $m=1$.) We find that $r_0$ and correspondingly the cosmological constant are restricted to
\be
1.585 \lesssim r_0 \lesssim 2.222 \sim r_{\rm cos}\,,\qquad\longleftrightarrow \qquad -0.313 \lesssim \Lambda \lesssim 0.0607\,,\label{lamcons}
\ee
in which case, the $R_{0, \rm min}$ lies in the region $2.675 \lesssim R_{0, \rm min} \lesssim 5.66$.

\begin{figure}[ht]
	\centering
	\includegraphics[width=0.6\textwidth]{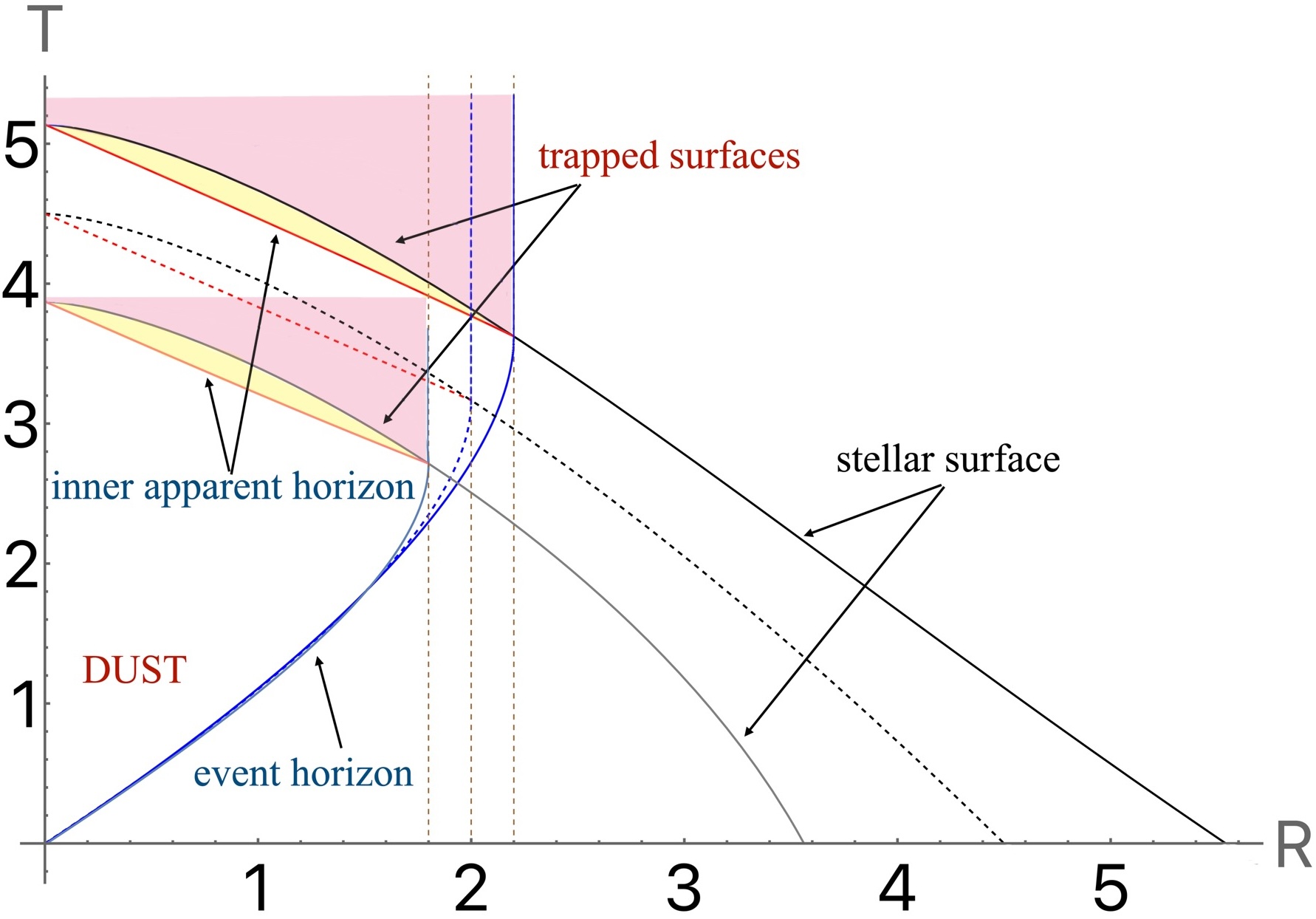}
	\caption{\small The $k=0$ collapsing of the $m=1$ black holes with a cosmological constant. The black and gray lines show the star's surface radius eq.~(\ref{tsch1}) for Schwarzschild-dS ($R_+=2.2$)  and AdS ($R_+=1.8$), respectively. The dotted line is the usual Schwarzschild black hole with $R_+=2$. The red and blue lines (dS), and pink and purple lines (AdS) show the apparent and event horizons, respectively, obtained by solving eqs.~(\ref{reh1}), (\ref{rapsch1}), and (\ref{tsch1}). For the dS case, $R_+ = 2.2$, $R_{0,\text{min}} = 5.54$, and for the AdS case, $R_+ = 1.8$, $R_{0,\text{min}} = 3.56$. Trapped surfaces are marked in yellow and pink for the interior and exterior regions of the star, respectively.}
	\label{f3}
\end{figure}

In Figure~\ref{f3}, using eq.~(\ref{tsch1}), we show the star's surface evolution for both Schwarzschild-dS (black line) and AdS (gray line) when $R_+ = 2.2$, $R_{0,\text{min}} = 5.54 < R_c \approx 5.94$ and $R_+ = 1.8$, $R_{0,\text{min}} = 3.56$, respectively, where the horizon starts formation from $R = 0$. Using eqs.~(\ref{rapsch1}) and~(\ref{reh1}), we solve for $R$ in terms of $R_{\text{ap}}$ and $R_{\text{eh}}$, and plot $T_{\text{ap}}(R)$ and $T_{\text{eh}}(R)$ for the same values of $R_+$ and $R_{0,\text{min}}$ (red and blue lines for dS, pink and purple lines for AdS).

To calculate the star's gravitational mass, density, and pressure, substitute eq. (\ref{sch1}) into eq. (\ref{den}) and (\ref{pre}) and returning $m$
\begin{align}\label{denc1}
 \rho(\tau) = \frac{3}{4\pi}\left(\frac{m}{R(\tau)^3} + \frac{\Lambda}{6}\right), \quad \quad p(\tau) = -\frac{\Lambda}{8\pi}.
\end{align}\label{pds}
Therfore we have 
\begin{align}
p=-\rho+\frac{3m}{4\pi R^3},
\end{align}
which shows that the equation of state for the stellar matter differs from the cosmological
constant equation of state, i.e. $p=-\rho$, by the second term of (\ref{pds}). Also eq. (\ref{denc1}) allows us to compare $m$ with the Misner-Sharp (MS) mass of the Schwarzschild-dS metric
\begin{align}\label{ms}
	m_{\text{MS}} = m + \frac{\Lambda}{6} R^3 = \frac{4\pi}{3} \rho R^3,
\end{align}
and as expected, the pressure in eq. (\ref{denc1}) is negative for $\Lambda > 0$.

Regarding the energy conditions, for $ \Lambda > 0 $, it is easy to check from eq.~(\ref{denc1}) that $ \rho \geq |p| $. This implies that the DEC, NEC, and WEC are all satisfied. For the SEC, $ \rho + 3p \geq 0 $; using eq.~(\ref{lambda}) and setting $ m = 1 $, we find
\begin{align}\label{sec}
	R \leq \frac{R_+}{(R_+ - 2)^{1/3}},
\end{align}
which provides a tighter upper bound on $ R_0 $. If we require the matter field to satisfy the SEC, then $ R_0 \leq R_+ / (R_+ - 2)^{1/3} $. For example, if $ R_+ = 2.2 $, satisfying the SEC requires $ R_0 < 3.76 $, which is less than $ R_{0,\text{min}}=5.54$. Therefore, the formation of a Schwarzschild-dS black hole requires a violation of the SEC inside the star. It should be pointed out that this calculation include the cosmological constant as the matter sector. If we exclude the contribution from the cosmological constant, all energy conditions are satisfied.

For $ \Lambda < 0 $, the NEC is satisfied, while the WEC, i.e., $ \rho \geq 0 $, requires
\begin{align}\label{weclambda}
	R > \left(\frac{2}{2 - R_+}\right)^{1/3} R_+,
\end{align}
which gives a lower bound on $ R_0 $. Comparing this with eq.~(\ref{r0ads}), we find that for black hole formation in the Schwarzschild-dS case, the WEC must be violated. For the DEC condition, which requires $ \rho - |p| > 0 $, we must have
\begin{align}\label{dec}
	R > (\frac{1}{2 - R_+})^{1/3} R_+.
\end{align}
This also gives a lower bound on $ R_0 $; in order to satisfy the DEC, we must have $ R_0 < R_{0,\text{min}} $. The SEC is always violated inside the star when $ R_+ < 2 $. Again, if we exclude the contributions from the cosmological constant, all energy conditions are satisfied.

\subsubsection{$k = 1$ case}

In this case, substituting eq. (\ref{sch1}) into eq. (\ref{tsrc}) gives
\begin{equation}\label{tsch1c}
T(R) = \int_{R_{\text{max}}}^R -\sqrt{\frac{3 R R_{\text{max}}}{(R - R_{\text{max}}) \left( -6 + R R_{\text{max}} \left( R + R_{\text{max}} \right) \Lambda \right)}} \, dR
\end{equation}
We require the collapse to start in the region $0 < R < R_{\text{max}}$. To avoid imaginary numbers in the square root, we need
$
-6 + R R_{\text{max}} \left( R + R_{\text{max}} \right) \Lambda  < 0,
$
which has two roots
\begin{equation}\label{root}
R = -\frac{R_{\text{max}}}{2} \pm \sqrt{\frac{R_{\text{max}}^2}{4} + \frac{6}{R_{\text{max}} \Lambda}}
\end{equation}
Solutions exist only for the positive branch and $\Lambda > 0$. Solving eq. (\ref{tsch1c}) gives 
\begin{align}\label{tcom}
	T(R) &= 
		\sqrt{\frac{R_{\text{max}} \left( 12 - R_{\text{max}}^3 \Lambda + \sqrt{R_{\text{max}}^3 \Lambda (24 + R_{\text{max}}^3 \Lambda)} \right)}
	{
		 (-6 + R R_{\text{max}} (R + R_{\text{max}}) \Lambda)}}
		\frac{1}{2\left(
		-3 \sqrt{R_{\text{max}}^3 \Lambda} + \sqrt{24 + R_{\text{max}}^3 \Lambda}
		\right)
	}\nonumber\\
	&\bigg[
	\frac{
		A \,\text{EllipticF}\left(\arcsin\left(\sqrt{\frac{B}{2}}\right), C\right)
	}{
		-3 + R_{\text{max}}^3 \Lambda
	}
	-
	\sqrt{-\frac{
			32 (-3 + R_{\text{max}}^3 \Lambda) (-6 + R R_{\text{max}} (R + R_{\text{max}}) \Lambda)
		}{
			R_{\text{max}} \Lambda
	}}
	\, \nonumber\\
	&\text{EllipticPi}\left(D, \arcsin\left(\sqrt{\frac{B}{2}}\right), C\right)
	\bigg],
\end{align}
where
\begin{align*}
	A &\equiv R_{\text{max}} \bigg( 
	6 \sqrt{24 + R_{\text{max}}^3 \Lambda} 
	+ \sqrt{R_{\text{max}}} \sqrt{\Lambda} 
	\big( 
	-30 R_{\text{max}} 
	+ R_{\text{max}}^4 \Lambda 
	+ R_{\text{max}}^{5/2} \sqrt{\Lambda} \sqrt{24 + R_{\text{max}}^3 \Lambda} 
	\\\nonumber
	&+ R \big( 
	12 
	+ 5 R_{\text{max}}^3 \Lambda 
	- 3 R_{\text{max}}^{3/2} \sqrt{\Lambda} \sqrt{24 + R_{\text{max}}^3 \Lambda}
	\big)
	\big)
	\bigg), 
\end{align*}
\begin{align*}
	B &\equiv 1 + \frac{
		-12 + R_{\text{max}}^2 (3 R + R_{\text{max}}) \Lambda
	}{
		\sqrt{R_{\text{max}}} (-R + R_{\text{max}}) \sqrt{\Lambda} \sqrt{24 + R_{\text{max}}^3 \Lambda}
	}, 
\end{align*}
\begin{align*}
	C\equiv \frac{
		2 \sqrt{R_{\text{max}}^3 \Lambda (24 + R_{\text{max}}^3 \Lambda)}
	}{
		-12 + R_{\text{max}}^3 \Lambda + \sqrt{R_{\text{max}}^3 \Lambda (24 + R_{\text{max}}^3 \Lambda)}},
\end{align*}
\begin{align*}
	D&\equiv  \left(
	\frac{1}{2} - \frac{3}{2} \sqrt{\frac{R_{\text{max}}^3 \Lambda}{24 + R_{\text{max}}^3 \Lambda}}
	\right)^{-1}.
\end{align*}
From the overall coefficient in eq. (\ref{tcom}), since we have $-6 + R R_{\text{max}} (R + R_{\text{max}}) \Lambda<0$, then we must have $ 12 - R_{\text{max}}^3 \Lambda + \sqrt{R_{\text{max}}^3 \Lambda (24 + R_{\text{max}}^3 \Lambda)}<0$, leading to 
\begin{align}
	\Lambda > \frac{3}{R_{\text{max}}^3}.
\end{align}
On the other hand the argumant of the Eliptic functions sould be real. So that we must have $B>0$ which gives
\begin{align}
	\Lambda < \frac{3}{R_{\text{max}}^3}.
\end{align}
Comparing the above equations we see that there is no solution and OS collapse model for the Schwarzschild-dS black hole when $k = 1$.

For the Schwarzschild-dS black hole, we can still model an OS collapse, in which the star begins collapsing from an initial radius $R_0 < R_{\text{max}}$. Substituting eq.~(\ref{sch1}) into eq.~(\ref{tsrc0}) yields the evolution of the star's surface, while the apparent and event horizons are determined from eqs.~(\ref{rehc}) and (\ref{rapc}), respectively. When choosing the parameters $R_+$, $R_{\text{max}}$, and $R_0$, care must be taken to ensure a physically consistent OS collapse. For example, with $R_{\text{max}} = 4$, $R_0 = 3.6$, and $R_+ = 2.2$, we obtain an OS collapse whose dynamics are similar to those shown in Fig.~\ref{f3}.
	
There exists a critical value $R_{\text{0,l}}$ that ensures the event horizon begins growing from a radius greater than zero. This value also limits the horizon formation time, as discussed in detail in Section~5.2. Substituting eq.~(\ref{sch1}) into eqs.~(\ref{tttc1}) and (\ref{r0c1}), and setting $m = 1$, for $R_+ = 2.2$ and $R_{\text{max}} = 4$, we obtain
	\begin{align}\label{schds}
		\Delta T_{\text{eh}} \approx 6.698, \quad R_{\text{0,l}} \approx 3.44.
	\end{align}

The forms of eqs.~(\ref{den}) and (\ref{pre}) remain unchanged for $k = 1$. Thus, the analysis of the energy conditions remains the same as in the $k = 0$ case.

\subsection{Reissner-Nordström Black Hole}

In this section, we consider the OS collapse for the RN black hole. The metric function is given by
\begin{align}\label{RN}
	f(R) = 1 - \frac{2}{R} + \frac{q^2}{R^2}.
\end{align}
The inner and outer horizons are
\begin{align}\label{RNh}
	R_{\pm} = 1 \pm \sqrt{1 - q^2}, \quad q^2 = R_{\pm}(2 - R_{\pm}),
\end{align}
where to have a real value for the charge $ q $, the outer horizon $ R_+ $ must be less than 2. It is known that the electromagnetic field modifies the Schwarzschild solution into the RN solution. During the collapse, as usual, we assume that the stellar particles do not carry electric charge, so their motion follows geodesics. Therefore, they experience the electromagnetic field only through changes in the spacetime metric, not through the Lorentz force. Early work on gravitational collapse with charge was studied in~\cite{R2}, considering self-gravitating charged dust with dynamical interior fields. Our model follows the OS framework, where a neutral dust star collapsing into a fixed exterior geometry.

\subsubsection{$ k = 0 $ case}

\begin{figure}[ht]
	\centering
	\includegraphics[width=0.6\textwidth]{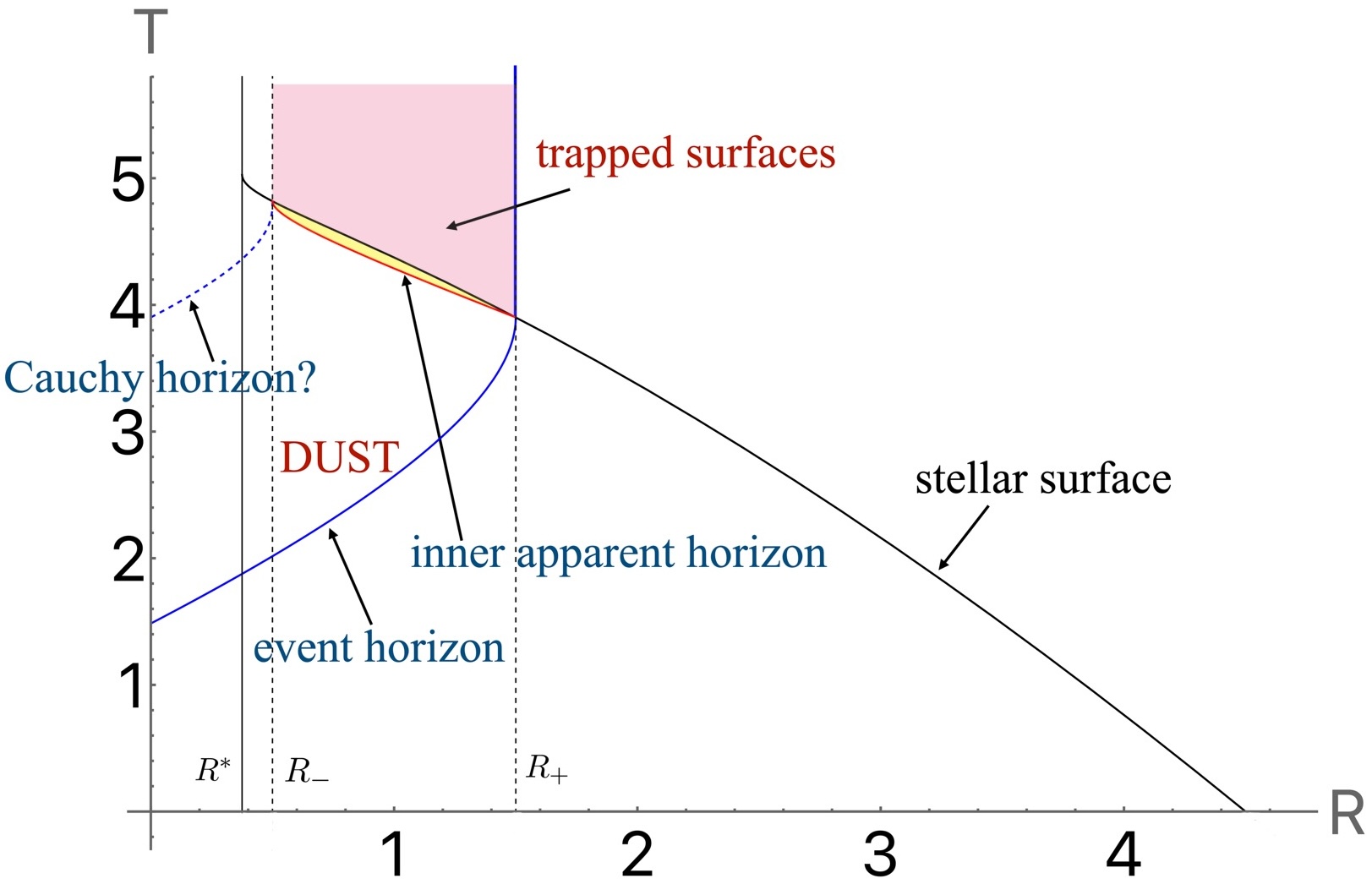}
	\caption{\small \textbf{$k = 0$, $R_0 = 4.5$:} The black line shows the star's surface from eq.~(\ref{RNs1}). The red and blue lines show the inner apparent and event horizons obtained from eqs.~(\ref{RNs1}), (\ref{RNrap1}), and (\ref{RNreh1}), respectively. The inner horizon evolution is also given by $T_{\text{ca}}(R)$ from eq. (\ref{RNreh2}), and represented by the blue dashed line. All three curves meet at $ R_+ = 1.5 $ and $ R_- = 0.5 $. Trapped surfaces are marked in yellow and pink for inside and outside areas of the star, respectively.}
	\label{f4}
\end{figure}

The surface radius of the collapsing star can be found by substituting eq.~(\ref{RN}) into (\ref{tsr}), which gives
\begin{align}\label{RNs1}
	T(R) = \frac{1}{3} \Big[ & \left(R_0 + R_+(2 - R_+)\right) \sqrt{2R_0 - R_+(2 - R_+)} \nonumber \\
	& - \left(R + R_+(2 - R_+)\right) \sqrt{2R - R_+ (2 - R_+)} \Big],
\end{align}
where we have expressed $ q $ in terms of $ R_+ $ using eq.~(\ref{RNh}). It should be noted that the reality condition for our calculations requires $1-f(R)>0$, or equivalently, the expression under the square root in the above equation must be positive, which gives 
\begin{align}\label{RNsin}
	R\ge R^*=\frac{q^2}{2}=\frac{R_+(2 - R_+)}{2}.
\end{align}
This shows that in the RN case, the star surface, or equivalently the test particle, will not reach $R=0$; instead, it goes to $R^*$. This occurs because the gravitational effective mass in this region is negative, resulting in repulsive gravity. In Figure~\ref{f4}, we show the evolution of the star's surface (black line) for $ R_+ = 1.5 $ and $ R_0 = 4 $, where eventually the star surface reaches $R^*=0.375$.

By substituting eq.~(\ref{RN}) into (\ref{rap}) and again expressing $ q $ in terms of $ R_+ $ from eq.~(\ref{RNh}), we obtain
\begin{equation}\label{RNrap1}
	R_{\text{ap}}(R) = \frac{R^2}{\sqrt{2R - R_+(2 - R_+)}}.
\end{equation}
Solving for $ R $ in terms of $ R_{\text{ap}} $, we plot $ T_{\text{ap}}(R) $ in Figure~\ref{f4} for $ R_+ = 1.5 $ and $ R_0 = 4 $ (red line). As shown, the inner apparent horizon crosses the surface at $ R_- = 0.5 $.  Similar to other discussed examples, the inner apparent horizon is always timelike for $ R_+ < 2 $, with
\begin{equation}
	y = \frac{R \left(3R - 2 R_+(2 - R_+) \right)}{\left(2R - R_+(2 - R_+)\right)^2}.
\end{equation}

For the event horizon, substituting eq.~(\ref{RN}) into (\ref{reh}) and using (\ref{RNh}), we find
\begin{align}\label{RNreh1}
	R_{\text{eh}}(R) = R \left( (1+R_+) - \sqrt{2R - R_+(2 - R_+)} \right),
\end{align}
with the integration constant fixed by the condition $ R_{\text{eh}}(R_+) = R_+ $. Solving for $ R $ in terms of $ R_{\text{eh}} $, we plot $ T_{\text{eh}}(R) $ in Figure~\ref{f4} for $ R_+ = 1.5 $ and $ R_0 = 4 $ (blue line). It should be noted that we can also fix the integration constant by the condition $ R_{\text{eh}}(R_-) = R_- $, where $R_- $ is the Cauchy horizon, and we have 
\begin{align}\label{RNreh2}
	R_{\text{ca}}(R) = R \left( (3-R_+) - \sqrt{2R - R_+(2 - R_+)} \right).
\end{align}
This line may represent the evolution of the Cauchy horizon inside the star. However, claiming this would be challenging. Again solving for $ R $ in terms of $ R_{\text{ca}} $, we plot $ T_{\text{ehca}}(R) $ in Figure~\ref{f4} for $ R_+ = 1.5 $ and $ R_0 = 4 $ (blue-dashed line).

In the extremal limit $ R_{\text{ext}} = R_{\pm} \to 1 $, the distance between the two horizons becomes smaller, and consequently, the length of the apparent horizon also becomes shorter. At the extremal point, there is no apparent horizon and therefore no trapped surfaces. 

\begin{figure}[ht]
	\centering
	\includegraphics[width=0.6\textwidth]{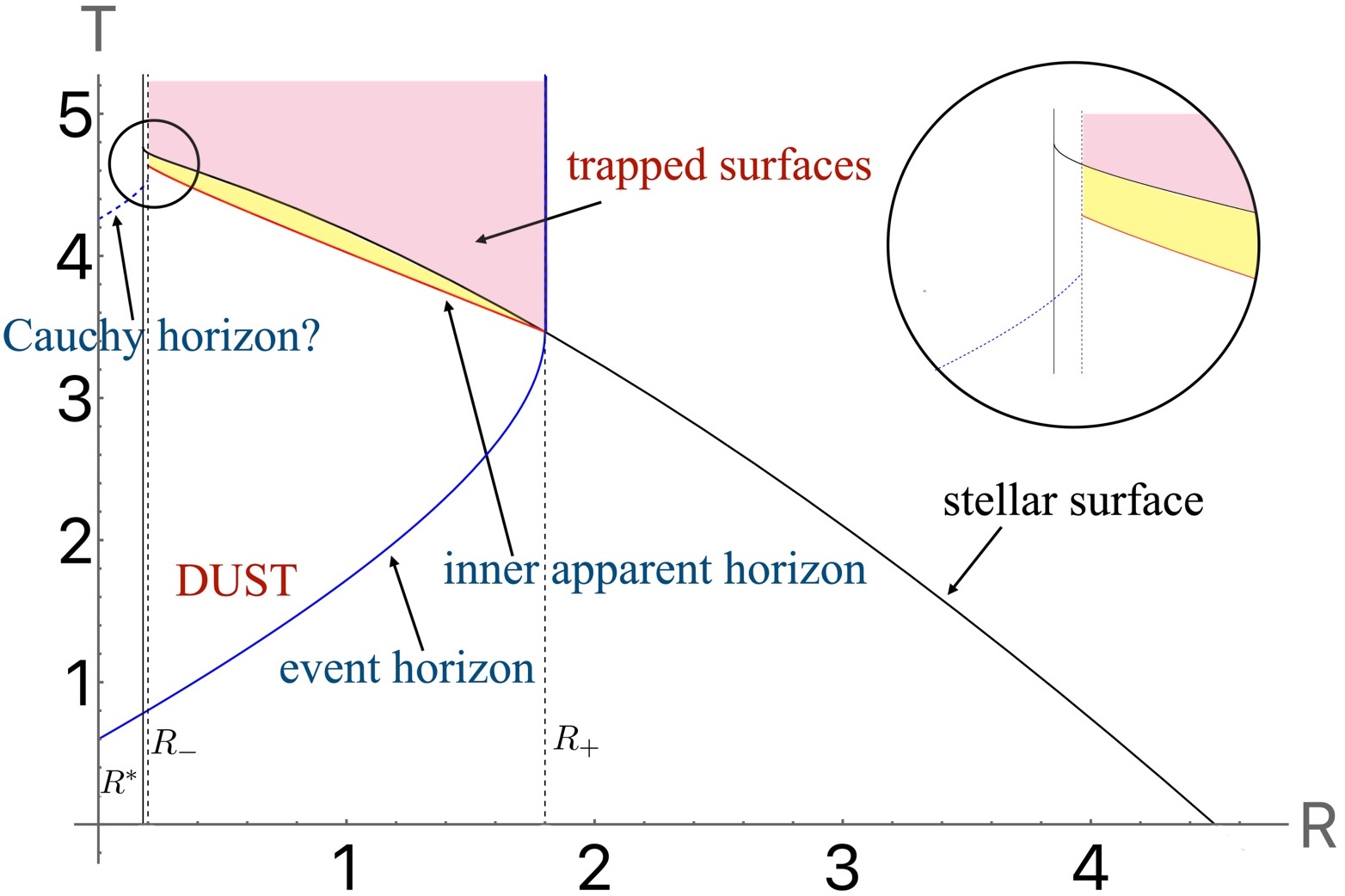}
	\caption{\small \textbf{$k = 0$, $R_0 = 4.5$:} The black, red, blue, and blue-dashed lines show $T(R)$, $T_{\text{eh}}(R)$, $T_{\text{ap}}(R)$, and $T_{\text{ca}}(R)$ obtained from eqs.~(\ref{RNs1}), (\ref{RNrap1}), (\ref{RNreh1}), and (\ref{RNreh2}), respectively. All three curves meet at $ R_+ = 1.8 $, but they do not meet at $ R_- = 0.2 $. Trapped surfaces are marked in yellow and pink for inside and outside areas of the star, respectively.}
	\label{f5}
\end{figure}

The inner apparent horizon is expected to begin at $R_+$ and end at $R_-$, which occurs when the star's surface crosses these two horizons. Thus, the surfaces trapped inside the black hole form a closed region in which 
$
T(R_{\pm})=T_{\text{ap}}(R_{\pm})
$. To achieve this $T(R_{-})=T_{\text{ap}}(R_{-})$, we need to have 
\begin{equation}\label{RNrap1c}
	0.75\le q^2 \le1 \hspace{0.5cm} \rightarrow \hspace{0.5cm} 1\le R_+\le 1.5, \hspace{0.5cm} 0.5\le R_-\le 1,
\end{equation}
which indicates a lower bound for the RN black hole charge as $q_{\text{min}}=0.75$ during the OS collapse process. Achieving $T(R_{-})=R_{\text{ap}}(R_{-})$ does not requires any conditions on parameters. Also, to have $T(R_{-})=T_{\text{ca}}(R_{-})$, we need to satisfy the same condition in eq. (\ref{RNrap1c}). In Figure~\ref{f5}, we plotted $T(R)$, $T_{\text{eh}}(R)$, $T_{\text{ap}}(R)$, and $T_{\text{ca}}(R)$ for $R_+=1.8$ and $R_0=4.5$. As it is obvious, the three lines $T(R)$, $T_{\text{ap}}(R)$, and $T_{\text{ca}}(R)$ do not intersect at $R=R_-$. This point shows that to have a consistent black hole formation in the RN case, we need to be closer to the extreme case where $q$ is large enough. The final point here is that both the WCCC \cite{3,4} and SCCC \cite{5} will be protected in this example since the star surface and thus the test particles on it, never reach the singularity at $R=0$.

Substituting eq.~(\ref{RN}) into (\ref{den}) and~(\ref{pre}), and restoring $ m $, the gravitational mass, density, and pressure of the star are
\begin{align}\label{denc1rn}
	\rho(\tau) = \frac{3}{4\pi} \left( \frac{m}{R(\tau)^3} - \frac{q^2}{2 R(\tau)^4} \right), \hspace{0.5cm}
	p(\tau) = -\frac{q^2}{8\pi R(\tau)^4},
\end{align}
where the pressure is negative. Therefore one can write 
\begin{align}\label{prn}
	\frac{\rho}{3}=p+(\frac{2 m^4 |p|^3}{\pi q^6})^{1/4},
\end{align}
which shows that the equation of state for the stellar matter differs from the radiation equation of state, i.e. $p=\rho/3$, by the second term of (\ref{prn}).

Regarding energy conditions, from eq.~(\ref{denc1rn}), satisfying both WEC and NEC requires
\begin{align}\label{wec}
	R > \frac{2 (R_+(2 - R_+))^2}{3},
\end{align}
which gives a lower bound on $ R_0 $. For example, when $ R_+ = 1.5 $, we approximately require $ R_0 > 0.375=R^*$. For the DEC, we require
\begin{align}\label{decrn}
	R > \frac{(R_+(2 - R_+))^2}{3},
\end{align}
which is again a lower bound on $ R_0 $, and smaller than the one required by the WEC and NEC. Finally, for the SEC, we require
\begin{align}\label{secrn}
	R > (R_+(2 - R_+))^2,
\end{align}
which gives a tighter lower bound on $ R_0 $. For example, if $ R_+ = 1.5 $, then to satisfy the SEC, we need $ R_0 > 0.56 $. It should be noted that in the extreme limit $ R_+ = 1 $, to satisfy the NEC, DEC, and SEC, we must have $ R_0 > \frac{2}{3} $, $ R_0 > \frac{1}{3} $, and $ R_0 > 1 $, respectively. These conditions are automatically satisfied because the lower bound on $ R_0 $, coming from the event horizon condition discussed in Section 5, is already larger than these values (see eq. (\ref{RNtr0})). 

\subsubsection{$ k = 1 $ case}

For $ k = 1 $, substituting eq.~(\ref{RN}) into (\ref{tsrc}) gives
\begin{align}\label{tsch1crn}
	T(R) = & \frac{R_{\text{max}} \sqrt{(R - R_{\text{max}}) \left(-2 R R_{\text{max}} - R_+ (R_+ - 2) (r + R_{\text{max}})\right)}}{R_+ (R_+ - 2) + 2 R_{\text{max}}} 
	\nonumber \\+ 
	& \frac{2 R_{\text{max}}^3}{\sqrt{\left(R_+ (2 - R_+) - 2 R_{\text{max}}\right)^3}} \, \text{ArcSinh}\left(\sqrt{\frac{(R - R_{\text{max}}) \left(R_+ - R_{\text{max}} - \frac{R_+^2}{2}\right)}{R_{\text{max}} \left(R_+ (2 - R_+) - R_{\text{max}}\right)}}\,\right),
\end{align}
where we substitute $ q $ in terms of $ R_+ $ using eq.~(\ref{RNh}). At $ R = 0 $, we get
\begin{align}\label{tsch1crn0}
	T(0) = \frac{\sqrt{R_+(R_+ - 2)} \, R_{\text{max}}^2}{(R_+ - 2) R_+ + 2 R_{\text{max}}} 
	+ 
	\frac{2 R_{\text{max}}^3 \, \text{ArcSinh}\left( \sqrt{ \frac{R_+ - R_{\text{max}} - \frac{R_+^2}{2}}{R_+ (2 - R_+) - R_{\text{max}}} } \right)}{\sqrt{(R_+ (R_+ - 2) + 2 R_{\text{max}})^3}},
\end{align}
which shows that for $ R_+ < 2 $ (real charge condition), the first term becomes imaginary. Therefore, $ T(R) $ does not reach $ R = 0 $. On the other hand, $ T(R_{\text{max}}) = 0 $, and the slope of $ T(R) $ is negative. Substituting $ R_{\text{max}} \to R_{\text{max}} - \epsilon^2 $ and expanding up to first order of $\epsilon$, we have
\begin{align}\label{tsch1crn0e}
	T(R_{\text{max}} - \epsilon^2) = - \sqrt{\frac{2 R_{\text{max}}^3}{R_{\text{max}} - \left(R_+ (2 - R_+)\right)^2}} \, \epsilon + \dots.
\end{align}
This implies $ T(R) < 0 $ in the region $ 0 \leq R \leq R_{\text{max}} $. Also, at $ R_+ = 2 $ (i.e., $ q = 0 $), eq.~(\ref{tsch1crn}) does not reduce to the Schwarzschild case and instead gives
\begin{align}
	T(R)|_{R_+ = 2} = \frac{\sqrt{R R_{\text{max}} (-R + R_{\text{max}})} - \sqrt{-R_{\text{max}}^3} \, \text{ArcSinh}\left(\sqrt{\frac{R - R_{\text{max}}}{R_{\text{max}}}}\right)}{\sqrt{2}}.
\end{align}
Therefore, there is no OS collapse for the RN black hole in the $k = 1$ case, where the star starts collapsing with zero velocity from $R_{\text{max}}$. However, one can still model an OS collapse in which the star begins collapsing from an initial radius $R_0 < R_{\text{max}}$. Substituting Eq.~(\ref{RN}) into Eq.~(\ref{tsrc0}), and using Eqs.~(\ref{rehc}) and (\ref{rapc}), yields the evolution of the star's surface, the apparent horizon, and the event horizon, respectively.

To have the surfaces trapped inside the black hole form a closed region, we require
$
T(R_{\pm}) = T_{\text{ap}}(R_{\pm}).
$, where the condition $T(R_{-}) = T_{\text{ap}}(R_{-})$ leads to
\begin{align}\label{RNrap1c1}
	&1 \le R_+ \le Y, \hspace{0.5cm} 1 - Y \le R_- \le 1, \notag \\ 
	&Y \equiv \frac{4}{3} - \frac{(1 - i \sqrt{3}) (-4 + 6 R_{\text{max}} - 6 R_{\text{max}}^2)}{3 \cdot 2^{2/3} X} + \frac{(1 + i \sqrt{3}) X}{6 \cdot 2^{1/3}}, \notag \\
	&X \equiv \left(16 - 36 R_{\text{max}} + 9 R_{\text{max}}^2 + 3 \sqrt{3} \sqrt{-48 R_{\text{max}}^2 + 136 R_{\text{max}}^3 - 157 R_{\text{max}}^4 + 96 R_{\text{max}}^5 - 32 R_{\text{max}}^6} \right)^{1/3}.
\end{align}
This implies a lower bound on the charge of the RN black hole. This result indicates that, for consistent black hole formation in the RN case, we must be close to the extremal limit, where the charge $q$ is sufficiently large. It should be noted that for different values of $R_{\text{max}}$, $Y$ takes real values in the range $1 < Y < 1.5$, and as $R_{\text{max}} \to \infty$, we have $Y \to 1.5$.

When choosing the parameters $R_+$, $R_{\text{max}}$, and $R_0$, care must be taken to ensure a physically consistent OS collapse, and the collapse process will be similar to that shown in Figs.~\ref{f4} and \ref{f5}. For example, with $R_{\text{max}}=5$, $R_{0,\text{min}} = 4$, and $R_+ = 1.4$, we obtain an OS collapse. There exists a critical value of $R_0$ that guarantees the event horizon begins growing from greater than zero, which also sets a limit on the horizon formation time; we will discuss this in detail in Section~5.2. The final point is that both the WCCC~\cite{3,4} and the SCCC~\cite{5} are preserved in the case $k=1$, since the star's surface—and thus any test particles on it—never reaches the singularity at $R=0$.

Since the forms of eqs.~(\ref{den}) and (\ref{pre}) are unchanged for $k = 1$, the analysis of the energy conditions proceeds similar to the $k = 0$ case.

\section{A conjecture on event horizon formation time}

In our $T/R$ graphs describing the OS collapsing of various cases, two curves are particularly noteworthy. One describes the stellar collapsing and the other describes the formation of the event horizon. They lead to a natural question: what is the characteristic time for the black hole formation in the OS collapsing. The $T(R)$ of the stellar surface is sensitively dependent on the initial $R_0$. On the hand, the quantity $\Delta T_{\rm eh}$ is independent of $R_0$. Thus the time needed for the event horizon formation $\Delta T_{\rm eh}$ would be a better time to describe the black hole collapsing. As we have seen in our earlier discussion, for the event horizon radius to evolve from zero to finally $R_+$, there must be a critical minimum initial $R_{0,\rm min}$ for the stellar surface. Thus $\Delta T_{\rm eh}$ also describes the minimum possible proper time of the stellar surface before it enters the black hole horizon $R_+$, where we have
\begin{align}
\Delta T_{\text{eh}}= T_{\text{eh}}(R_+)-T_{\text{eh}}(0)=T(R_+)-T(R_{0,\text{min}}),
\end{align}
and we will show in eq. (\ref{r0}) that $R_{0,\text{min}}$ can be obtained in terms of $R_+$.

\subsection{$k = 0$ case}

This situation is easy to analyze. It follows from eq.~(\ref{reh}), the solution, satisfying the boundary condition $R_{\text{eh}}(R_+) = R_+$, is given by
\begin{align}\label{rreh}
	R_{\text{eh}}(R) = R \left(1 - \int_{R_+}^R \frac{dR}{R \sqrt{1 - f(R)}}\right).
\end{align}
Requiring $ R_{\text{eh}} = 0 $ (equivalent to $ T_{\text{eh}} = 0 $) gives the minimum value of $ R_0 $, denoted as $ R_{0,\text{min}} $. If $R_0<R_{0,\text{min}}$, then the event horizon would start forming at a negative time, which is unphysical. It is given by
\begin{align}\label{r0}
	1 = \int_{R_+}^{R_{0,\text{min}}} \frac{dR}{R \sqrt{1 - f(R)}}.
\end{align}
The characteristic minimum black hole formation time, measured as the proper time of stellar surface, is then given by
\begin{align}\label{ttt}
	\Delta T_{\rm eh} = \int_{R_0,\rm min}^{R_+} -\frac{dR}{\sqrt{1 - f(R)}}.
\end{align}
For the Schwarzschild black hole (see  appendix A), we have concluded that
\begin{align}\label{schrt}
	\Delta T_{\text{eh(sch)}} = \frac{19}{6}m = \frac{19}{12} R_{+(\text{sch})}, \quad R_{0,\text{min(sch)}} = \frac{9}{2}m = \frac{9}{4} R_{+(\text{sch})}.
\end{align}
It is natural to compare this $\Delta T_{\rm eh}$ quantities among different black holes for fixed mass. Naively, one might expect that when mass is fixed, it would take less time to form a larger black hole, but our results for the collapsing of the Schwarzschild-AdS/dS black holes indicate the opposite. For sufficiently small cosmological constant, a perturbative analytical result was given in the previous section. In particular, black hole radius is smaller for negative cosmological constant and larger for positive cosmological constant. The collapsing time $\Delta T_{\rm eh}$, given in \eqref{deltaTlinearep}, shows that the $\Delta T_{\rm eh}$ is shorter for smaller $R_+$ and larger for larger $R_+$. Our numerical analysis indicates this is generally true for non-infinitesimal cosmological constant, but restricted by \eqref{lamcons} for the Schwarzschild-AdS/dS black holes.

Of course, one may argue that it only makes sense to compare $\Delta T_{\rm eh}$ for different black holes in the same asymptotic geometry, such as Minkowski or (A)dS spacetimes. So we consider the OS collapse for the RN black hole. In this case, we have
\begin{align}\label{RNtr0}
	\Delta T_{\text{eh(RN)}} &= \frac{5 m}{3}+\frac{3}{2}\sqrt{m^2-q^2}= \frac{5 m}{3}+\frac{3}{2} (R_{+(\text{RN})} - m), \nonumber \\
	R_{0,\text{min(RN)}} &= \frac{5m}{2} + 2 \sqrt{m^2 - q^2} = \frac{5m}{2} + 2 (R_{+(\text{RN})} - m),
\end{align}
where we substituted $ q $ in terms of $ R_+ =m + \sqrt{m^2-q^2}$. Again we see that $\Delta T_{\text{eh(RN)}}$ is positively proportional to the $R_+$. In other words, for fixed mass $m$, larger black holes require longer $\Delta T_{\text{eh(RN)}}$.

For asymptotically flat black holes, the Schwarzschild black hole is the largest \cite{Lu:2019zxb,Yang:2019zcn}, thus we expect the characteristic black hole formation time $\Delta T_{\rm eh}$ for the Schwarzschild black hole sets an upper bound for the formation time of black holes with given mass, namely
\be\label{con1}
\Delta T_{\rm eh} \le \fft{19}{6} m\,.
\ee
We are unable to prove this in general. However, we can prove that the Schwarzschild black hole set an upper bound also for $R_{0,\rm min}$, as given in \eqref{schrt}. This is because for general black holes, we have $f=1-2m(r)/r$, where $m(r)$ is a monotonically increasing function with $m(\infty)=m$ and $m(r_+)=r_+/2$. Together with the fact that $r_+$ is largest for the Schwarzschild black hole, it follow from the constraint \eqref{r0} that $R_{0,\rm min}$ is the largest for the Schwarzschild black hole with fixed mass.

\subsection{$k = 1$ case}

For $ k = 1 $, imposing $ R_{\text{eh}}(R_+) = R_+ $ in eq.~(\ref{rehc}) and solving for $ R_{\text{eh}} $ gives
\begin{align}\label{rrehc}
	R_{\text{eh}}(R) = R \left(1 - \int_{R_+}^R \sqrt{\frac{f(R_{\text{max}})}{f(R_{\text{max}}) - f(R)}} \frac{dR}{R}\right).
\end{align}
Setting $ R_{\text{eh}} = 0 $ gives the lower bound $ R_{\text{max,l}} $
\begin{align}\label{r0c}
	1 = \int_{R_+}^{R_{\text{max,l}}} \sqrt{\frac{f(R_{\text{max,l}})}{f(R_{\text{max,l}}) - f(R)}} \frac{dR}{R}.
\end{align}
The characteristic minimum black hole formation time, measured as the proper time of the stellar surface, is then given by
\begin{align}\label{tttc}
	\Delta T_{\rm eh} = \int_{R_{\text{max,l}}}^{R_+} -\frac{dR}{\sqrt{f(R_{\text{max}}) - f(R)}}.
\end{align}
For the Schwarzschild black hole, substituting eq.~(\ref{sch1}) into eqs.~(\ref{tttc1}) and (\ref{r0c1}) yields
\begin{align}\label{schrtc}
	\Delta T_{\text{eh(sch)}} \approx 4.536m = 2.268 R_{+(\text{sch})}, \quad R_{\text{max,l(sch)}} \approx 3.171m = 1.585 R_{+(\text{sch})}.
\end{align}
As discussed earlier, there is no OS collapse for the other examples considered, where $R_0=R_{\text{max}}$.

Same as for $ k = 0 $, for $ k = 1 $ and asymptotically flat black holes, the Schwarzschild black hole is the largest \cite{Lu:2019zxb,Yang:2019zcn}. Thus, we expect that the $\Delta T_{\rm eh}$ for the Schwarzschild black hole sets an upper bound for the formation time of black holes with the same mass, as
\begin{align}\label{conc1}
	\Delta T_{\text{eh}} \leq \Delta T_{\text{eh(sch)}} = 2.268 R_{+(\text{sch})}.
\end{align}
Comparing eqs.~(\ref{con1}) and (\ref{conc1}), we see that upper bound for the formation time of black holes is smaller for $ k = 0 $ than for $ k = 1 $.

For the case where the star starts collapsing from $R_0 < R_{\text{max}}$ with an initial velocity, setting $R_{\text{eh}} = 0$, eq.~(\ref{r0c}) takes the following form and gives the lower bound $R_{\text{max,l}}$
	\begin{align}\label{r0c1}
		1 = \int_{R_+}^{R_{\text{0,l}}} \sqrt{\frac{f(R_{\text{max}})}{f(R_{\text{max}}) - f(R)}} \frac{dR}{R}.
	\end{align}
	When $R_0 = R_{\text{max}}$, $R_{\text{max,l}}$ can be obtained as a function of only $R_+$ from Eq.~(\ref{r0c}). For $R_0 < R_{\text{max}}$, $R_{\text{0,l}}$ is a function of both $R_+$ and $R_{\text{max}}$. The characteristic minimum black hole formation time is then given by
	\begin{align}\label{tttc1}
		\Delta T_{\rm eh} = \int_{R_{\text{0,l}}}^{R_+} -\frac{dR}{\sqrt{f(R_{\text{max}}) - f(R)}}.
	\end{align}
	For the Schwarzschild black hole, substituting eq.~(\ref{sch0}) into eqs.~(\ref{tttc1}) and (\ref{r0c1}) yields
	\begin{align}\label{schrtc1}
		R_{\text{max}}=5 \,\,\rightarrow \,\, \Delta T_{\text{eh(sch)}} &\approx 3.894\,m = 1.947  R_{+(\text{sch})}, \quad R_{\text{0,l(sch)}} \approx 3.95\, m = 1.975 R_{+(\text{sch})},\nn\\
		R_{\text{max}}=6 \,\,\rightarrow \,\, \Delta T_{\text{eh(sch)}} &\approx 3.755\,m = 1.877  R_{+(\text{sch})}, \quad R_{\text{0,l(sch)}} \approx 4.09\, m = 2.045 R_{+(\text{sch})}.
	\end{align}
	The above results show that as $R_{\text{max}}$ decreases toward $R_{\text{max,l}} = 3.171\,m$, $R_{\text{0,l}}$ also decreases toward $3.171\,m$, while $\Delta T_{\text{eh(sch)}}$ increases. This supports the validity of the conjecture proposed in eq.~(\ref{conc1}).
	
	Now, let us examine the validity of this conjecture in the RN case with $R_0 < R_{\text{max}}$. Substituting eq.~(\ref{RN}) into eqs.~(\ref{tttc1}) and (\ref{r0c1}) and setting $m=1$, for $R_+ = 1.9$ we obtain
	\begin{align}\label{RNrtc1}
		R_{\text{max}}=4 \,\,\rightarrow \,\, \Delta T_{\text{eh(RN)}} &\approx 3.912, \quad R_{\text{0,l(RN)}} \approx 3.577, \nn\\
		R_{\text{max}}=5 \,\,\rightarrow \,\, \Delta T_{\text{eh(RN)}} &\approx 3.678, \quad R_{\text{0,l(RN)}} \approx 3.8,
	\end{align} 
	and for $R_+ = 1.8$ we obtain
	\begin{align}\label{RNrtc2}
		R_{\text{max}}=4 \,\,\rightarrow \,\, \Delta T_{\text{eh(RN)}} &\approx 3.702, \quad R_{\text{0,l(RN)}} \approx 3.462, \nn\\
		R_{\text{max}}=5 \,\,\rightarrow \,\, \Delta T_{\text{eh(RN)}} &\approx 3.488, \quad R_{\text{0,l(RN)}} \approx 3.656.
	\end{align} 
	These numerical results clearly confirm the validity of the conjecture proposed in eq.~(\ref{conc1}).

As in the case $k=0$, for asymptotically AdS/dS spacetime and when $k=1$, $\Delta T_{\rm eh}$ is longer for larger $R_+$, as shown in eq.~(\ref{schds}).

\section{Conclusion}

In this work, we have extended the OS model of gravitational collapse within the framework of standard Einstein gravity. We consider a collapsing star whose interior is described by a homogeneous and isotropic FLRW geometry ($k=0$ or $k=1$), matched smoothly to a static, spherically symmetric exterior spacetime that is a solution of the Einstein equations—such as Schwarzschild, RN, and Schwarzschild-AdS/dS. Our analysis is fully embedded in general relativity; the exterior is not completely an arbitrary ansatz, but a physically motivated black hole geometry sourced by standard matter content (e.g., electromagnetic field or cosmological constant). To facilitate a smooth junction, we employ coordinate systems adapted to freely falling observers: PG coordinates for $k=0$ and a Novikov-like coordinate system for $k=1$, which eliminate horizon coordinate singularities and enable direct matching at the star's surface.

By matching the interior FLRW geometry with the generalized exterior metric, we derived the equation governing the motion of the stellar surface $R$ as: 
$
\dot{R} = \sqrt{1 - f(R)} \quad (k=0), \quad \dot{R} = \sqrt{f(R_{\text{max}}) - f(R)} \quad (k=1).
$
We also obtained expressions for the energy density and pressure of the matter inside the star and confirmed that the net radial pressure at the surface vanishes, ensuring that particles follow geodesics during the collapse. Furthermore, we verified that the junction conditions are satisfied by computing the extrinsic curvature on both sides of the stellar surface. The continuity of both the metric and the extrinsic curvature confirms that the two spacetimes can be joined smoothly without introducing any thin-shell matter at the boundary.

In Section 3, we studied how black holes form dynamically during the gravitational collapse of a star. We focused on two main aspects: the formation of horizons and the behavior of matter during collapse. First, we analyzed the formation of event and apparent horizons during the collapse. In both the flat ($ k=0 $) and closed ($ k=1 $) cases, the inner apparent horizon (inside the star) appears when the event horizon intersects the star surface and grows inward as the collapse continues. On the other hand, the event horizon starts from zero, $ R = 0 $, and expands outward, eventually matching the outer black hole horizon (outside the star). Second, we checked whether the collapsing matter satisfies the standard energy conditions — such as the WEC, NEC, DEC, SEC. These conditions ensure that the matter behaves in a physically reasonable way — for instance, that the energy density is nonnegative.

In the next sections, we applied our generalized OS collapse model to explicit examples: Schwarzschild–AdS/ds and RN black holes. For the Schwarzschild-dS black hole, OS collapse is possible in the $k=0$ case provided the collapse begins inside the cosmological horizon, which imposes an upper bound on $|\Lambda|$ for a fixed black hole mass. In the $k=1$ case, consistent OS collapse is not possible if the star begins collapsing from rest at $R_0 = R_{\text{max}}$ (zero initial velocity). However, as it is shown, a physically consistent collapse can be achieved when the star starts from an initial radius $R_0 < R_{\text{max}}$ with non-zero initial velocity. Similarly, for the Schwarzschild-AdS case with fixed mass, OS collapse is viable only for sufficiently small $|\Lambda|$. To provide more examples, we extended our analysis of gravitational collapse to the RN black hole. This case introduces new features such as the presence of two horizons — an outer event horizon $ R_+ $ and an inner Cauchy horizon $ R_- $ — and a repulsive gravitational effect near the center due to the charge term. For the spatially flat ($ k=0 $) case, we found that the star collapses smoothly, but unlike in the Schwarzschild case, it does not reach $ R = 0 $. Instead, it reaches a minimum radius $ R^* $. This happens because gravity becomes repulsive in this region, and it can be infinitely repulsive as the geodesic moves to the center. We derived explicit expressions for the evolution of the stellar surface, apparent horizon, and event horizon, showing how the inner apparent horizon connects the two horizons and forms a closed region of trapped surfaces, where the condition $ q^2 \geq \frac{3}{4} $ should be satisfied. We also analyzed the extremal limit, where the two horizons merge. In this case, the apparent horizon disappears, and the collapse dynamics resemble those of a Schwarzschild black hole, where the star surface reaches $ R^* = \frac{1}{2} $. Regarding energy conditions, we verified that the Weak, Null, Dominant, and Strong Energy Conditions hold under certain constraints on the initial radius $ R_0 $. However, for the spatially closed ($ k=1 $) case, we found that no consistent OS collapse solution exists. Attempts to derive a real-valued collapse trajectory lead to imaginary time evolution, indicating that RN collapse cannot proceed in a closed universe under these assumptions.

In the appendix, we also considered the Schwarzschild black hole, both the spatially flat ($ k=0 $) and closed ($ k=1 $) cases show consistent gravitational collapse leading to black hole formation. While the collapse is slightly faster in the $ k=0 $ case, both scenarios result in the formation of well-defined event and apparent horizons, and all energy conditions are satisfied. The results in this section have been used in section. 5, where we proposed a conjecture on horizon formation time. 

Our analysis improves previous studies in several key respects. Unlike Ref.~\cite{r2}, we show that for $k=1$ with momentarily static initial data ($R_0 = R_{\text{max}}$), no physically consistent OS collapse exists in the Schwarzschild-de Sitter spacetime. However, if the collapse begins from $R_0 < R_{\text{max}}$ with non-zero initial velocity, a consistent solution can be achieved. Moreover, we demonstrate that there exists a lower bound $R_{0,\text{min}}$ such that for $R_0 < R_{0,\text{min}}$, the event horizon would begin forming at a negative exterior time, which is unphysical. 
Also comparing with Ref.~\cite{R2}, who studied self-gravitating charged dust with dynamical interior electromagnetic fields, our RN treatment adheres to the OS framework, where a neutral dust star collapses into a fixed exterior geometry. Furthermore, using PG coordinate for $k=0$ and a novel Novikov-like system for $k=1$—enables the systematic computation of horizon evolution, event horizon formation time $\Delta T_{\text{eh}}$, the minimum initial radius $R_{0,\text{min}}$, and energy conditions within a single formalism. These results demonstrate that while the OS collapse model works well in Schwarzschild spacetime, its generalization to more complex backgrounds introduces new complexities and constraints. These findings highlight the sensitivity of gravitational collapse to the global structure of spacetime and the nature of the surrounding matter fields.

In section 5, we identified two key features of black hole formation during gravitational collapse, for both spatially flat ($ k=0 $) and closed ($ k=1 $) geometries:
\begin{enumerate}
	\item For a black hole with fixed mass and external parameters (e.g., $\Lambda$ or $q$), there exists a critical minimum initial radius $R_{0,\text{min}}$ such that collapse must begin from $R_0 > R_{0,\text{min}}$ to ensure physical (non-negative) horizon formation time.
	\item The total time for the event horizon to form — denoted $ \Delta T_{\text{eh}} $ — depends only on the properties of the black hole. Importantly, we found that $ \Delta T_{\text{eh}} $ is a function of the black hole's event horizon radius $ R_+ $, and not of the initial radius $ R_0 $.
\end{enumerate}
These results led us to propose a conjecture on the collapse time inequality. It states that for the given mass $M$ and the cosmological constant, the collapsing time, from the critical $R_{0,\rm min}$ to the eventual formation of the stable $R_{\rm eh}=R_+$, is the longest for the Schwarzschild black hole. For asymptotically Minkowski spacetimes, this inequality states
\begin{equation}
	\Delta T_{\text{eh}} \le \Delta T_{\text{eh(sch)}} = \fft{19}6 M.
\end{equation}
This inequality is counterintuitive since for a given mass, the Schwarzschild black hole has the largest size \cite{Lu:2019zxb,Yang:2019zcn}, and one would naively expect that it would take less time for the dust star to collapse into a larger black hole. The puzzle may be explained by the fact that our  $\Delta T_{\text{eh}}$ is effectively calculated as the growing time of the event horizon from $0$ to $R_+$.

This work introduces a new perspective on black hole formation, offering both a practical tool for predicting horizon formation times and a potential avenue for a deeper understanding of gravitational collapse in general relativity.

\section*{Acknowledgement}

We are grateful to Run-Qiu Yang for useful discussions. H.K.~and H.L.~are supported in part by the National Natural Science Foundation of China (NSFC) grants No.~12375052, No.~11935009 and No.~W2533015, as well as by the Tianjin University Self-Innovation Fund Extreme Basic Research Project Grant No.~2025XJ21-0007.

\section*{Appendix}

\begin{appendix}

\renewcommand{\theequation}{A.\arabic{equation}}  
\setcounter{equation}{0} 
	
\section{Schwarzschild black hole}
	
	The Schwarzschild black hole is given by
	\begin{equation}\label{sch0}
		f(R) = 1 - \frac{2}{R},
	\end{equation}
	where we set $m = 1$ for simplicity (equivalently, one can define a dimensionless parameter by rescaling $R \to R/m$). The event horizon is located at $R_+ = 2$.
	
	\subsection{$k = 0$ Case}
	
	To determine the evolution of the star's surface radius, substituting eq. (\ref{sch0}) into eq. (\ref{tsr}) gives
	\begin{equation}\label{tsch}
		T(R) = \frac{\sqrt{2}}{3} \left(R_0^{3/2} - R^{3/2}\right),
	\end{equation}
	where on the horizon $R_+ = 2$, we find $T(R_+) = (\sqrt{2} R_0^{3/2} - 4)/3$. This equation describes the evolution of the star's radius from an arbitrary initial value $R = R_0>R_+$ to $R = 0$ over a finite time $T(0)$.
	
	For the apparent horizon, substituting eq. (\ref{sch0}) into eq. (\ref{rap}) yields
	\begin{equation}\label{rapsch}
		R_{\text{ap}}(R) = \frac{R^{3/2}}{\sqrt{2}}.
	\end{equation}
	Solving for $R$ in terms of $R_{\text{ap}}$ and substituting into eq. (\ref{tsch}), we obtain the time evolution of the apparent horizon as
	\begin{equation}\label{tapsch}
		T_{\text{ap}}(R) = \frac{\sqrt{2}}{3} \left(R_0^{3/2} - {\sqrt{2}} R\right),
	\end{equation}
	where $T(R_+) = T_{\text{ap}}(R_+)$. The apparent horizon begins growing from $R = R_+$ and reaches $R = 0$ in alignment with the star's surface radius within a finite time $T_{\text{ap}}(0)$. To determine whether the apparent horizon is timelike or spacelike, substituting eq. (\ref{sch0}) into eq. (\ref{n2}) leads to $x = -3/2$ and $n^2 = 3/4 > 0$. Thus, during gravitational collapse, the apparent horizon is a timelike surface.
	
	For the event horizon's evolution, substituting eq. (\ref{sch0}) into eq. (\ref{reh}) gives
	\begin{equation}\label{rehsch}
		R_{\text{eh}}(R) = -\sqrt{2} R^{3/2} + 3R,
	\end{equation}
	where the integration constant is determined from the condition $R_{\text{eh}}(R_+) = R_+$. Solving this equation for $R$ in terms of $R_{\text{eh}}$ yields three roots, but only one lies within the star's region. Thus, we obtain
	\begin{equation}\label{tehsch}
		T_{\text{eh}}(R) = \frac{\sqrt{2}}{3} \left(R_0^{3/2} - \frac{1}{2\sqrt{2}} \left(3 + \frac{9 - 4R}{W} + W\right)^{3/2}\right),
	\end{equation}
	where:
	\begin{equation}\label{w}
		W \equiv \left(27 + 2R(-9 + R) + 2\sqrt{(-2 + R)R^3}\right)^{1/3}.
	\end{equation}
	On the horizon, we have $T(R_+) = T_{\text{ap}}(R_+) = T_{\text{eh}}(R_+)$. The event horizon begins growing during the collapse from $R = R_i$ when $T_{\text{eh}} = 0$ and eventually crosses both the apparent horizon and the star's surface at $R = R_+$ within a finite time $T(R_+)$.
	
\begin{figure}[ht]
		\centering
		\includegraphics[width=0.6\textwidth]{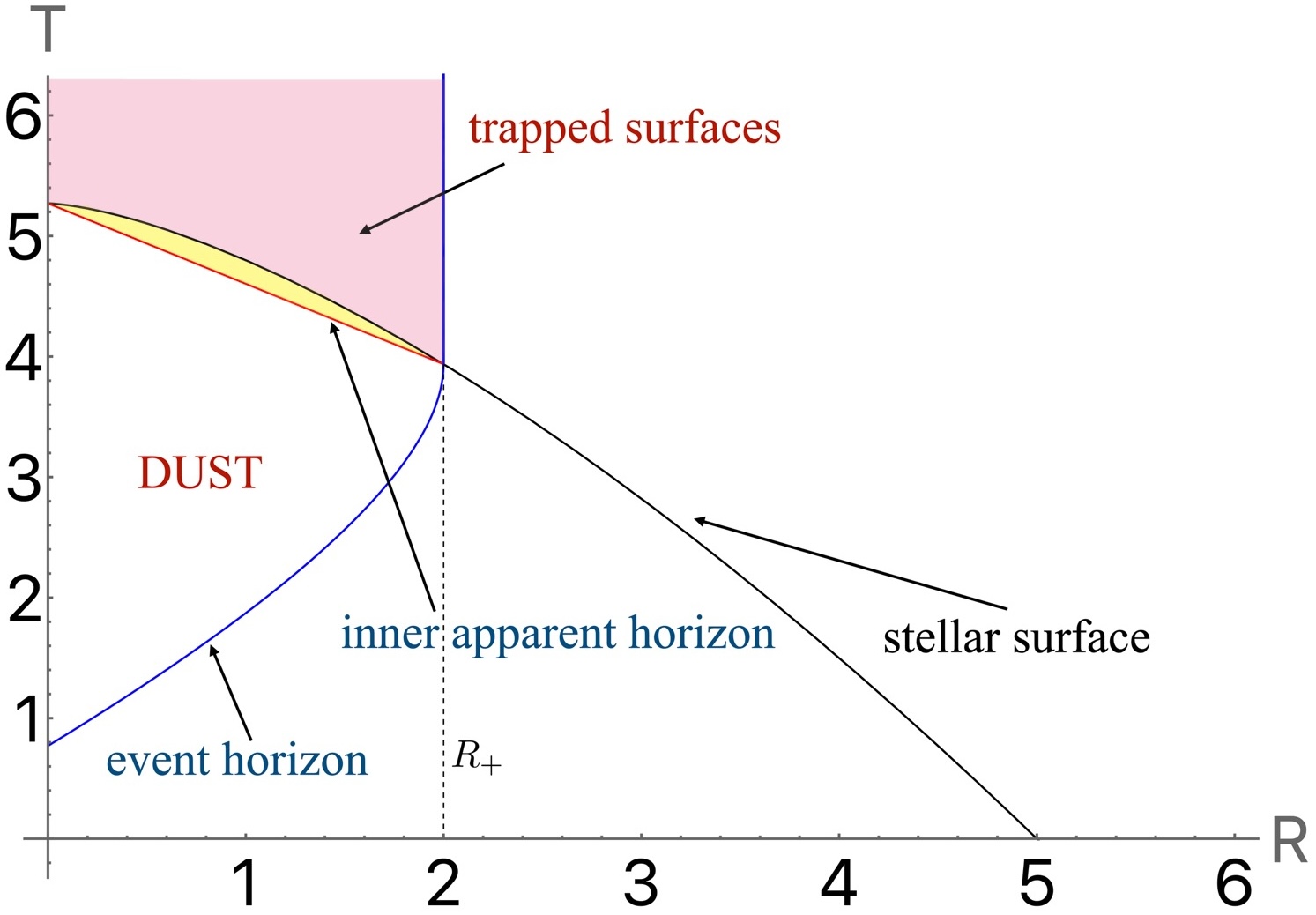}
		\caption{\small The collapsing of the Schwarzschild of $m=1$. We consider $k = 0$, $R_0 = 5$. The black line represents the star's surface radius (\ref{tsch}), while the red and blue lines correspond to the interior apparent and event horizons, respectively, given by eqs. (\ref{tapsch}) and (\ref{tehsch}). All three curves intersect at $T = (-4 + 5\sqrt{10})/3$ and $R_+ = 2$. Trapped surfaces are marked in yellow and pink for inside and outside areas of the star, respectively.}
		\label{f1}
	\end{figure}

	In Fig.~\ref{f1}, we plot the evolution of the star's surface radius, apparent horizon, and event horizon given by eqs. (\ref{tsch}), (\ref{tapsch}), and (\ref{tehsch}) for the starting initial $R_0 = 5$. The star begins collapsing from $R = 5$. At certain critical time, the event horizon develops and starts growing and reaches the star's surface at $R = 2$. At this point, the interior apparent horizon also emerges from $R = 2$ and shrinks to $R = 0$ in alignment with the star's surface radius. Recall that initially star interior can have both ingoing and outgoing geodesics. When the interior apparent horizon forms, interior matter inside the apparent horizon can still have both outgoing and ingoing geodesics, but the interior matter outside the apparent horizon 
	(the yellow region in Fig.~\ref{f1}) can have only ingoing geodesics. The interior apparent horizon thus shrinks with time.
	
	Let us now calculate the gravitational mass, density, and pressure of the collapsing star. Substituting eq. (\ref{sch0}) into eqs. (\ref{den}) and (\ref{pre}) and recalling $m$, we find
	\begin{equation}\label{densch}
		\rho(\tau) = \frac{3}{4\pi}\frac{m}{R(\tau)^3}, \quad \quad p(\tau) = 0.
	\end{equation}
	The constancy of the gravitational mass $m$ reflects energy conservation during the collapse, as no mass is lost to radiation or external forces in this idealized scenario. The time-dependent density $\rho(\tau) > 0$ increases monotonically as the star contracts, diverging to infinity as $R(\tau) \to 0$. The result of zero pressure ($p(\tau) = 0$) indicates the hidden assumption in the OS formalism that there are no internal forces, such as gas or radiation pressure to oppose gravity. Note that although the pressure is the same on the interior of the star surface as its exterior, the energy density $\rho$ is not. It is thus necessarily to check the energy conditions of the collapsing matter. It follows from (\ref{densch}) that the matter field satisfies all energy conditions.
	
	\subsection{$k = 1$ Case}
	
\begin{figure}[ht]
		\centering
		\includegraphics[width=0.6\textwidth]{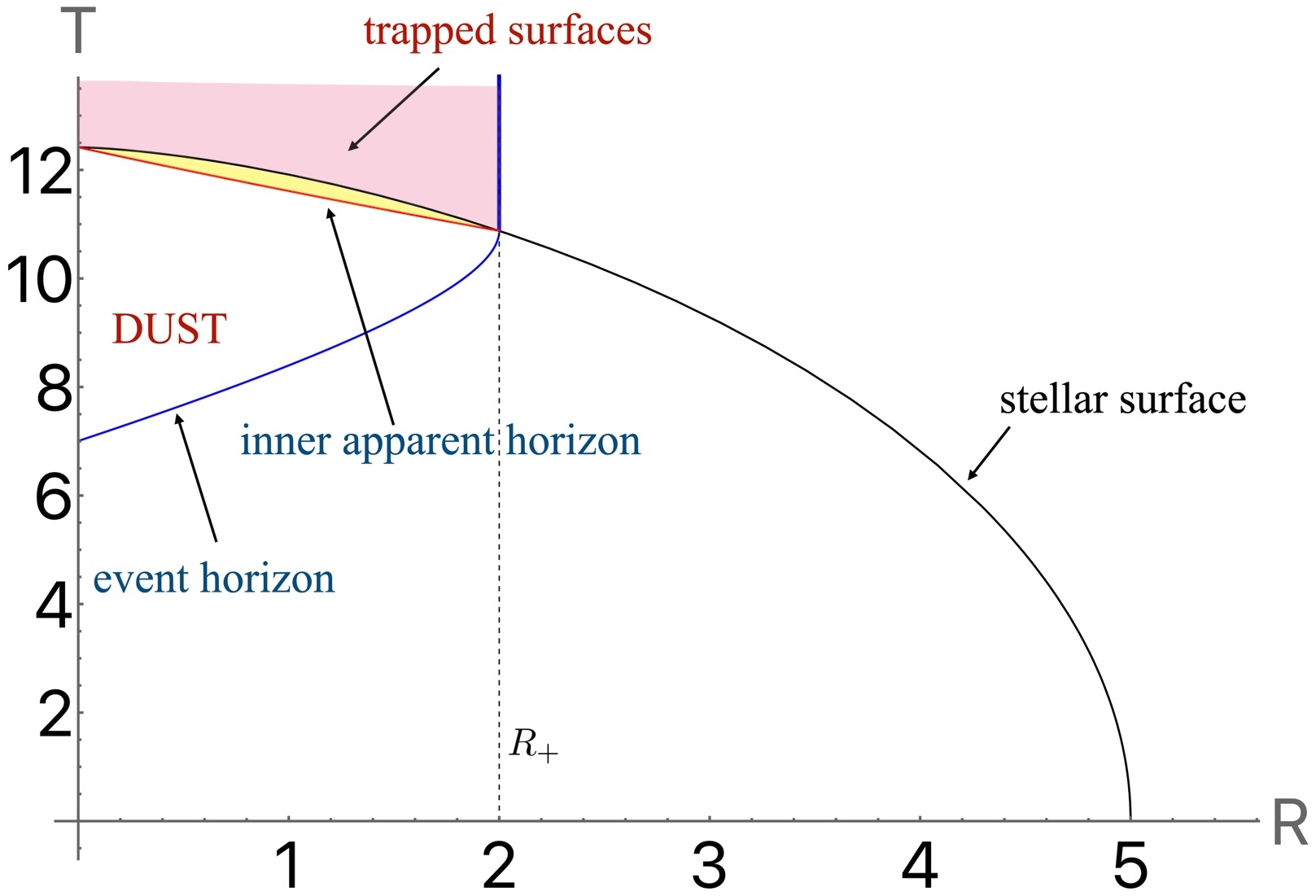}
		\caption{The collapsing of the Schwarzschild of $m=1$, with $k = 1$, $R_{\text{max}} = 5$. The black line represents the star's surface radius (\ref{tschc}), while the red and blue lines correspond to the apparent and event horizons, respectively, obtained by numerically solving eqs. (\ref{rapschc}), (\ref{rehschc}), and (\ref{tschc}). All three curves intersect at $R_+ = 2$. Trapped surfaces are marked in yellow and pink for inside and outside areas of the star, respectively.}
		\label{f2}
	\end{figure}

	To calculate the star's surface radius evolution for $k = 1$, we substitute eq. (\ref{sch0}) into eq. (\ref{tsrc}):
	\begin{equation}\label{tschc}
		T(R) = \sqrt{\frac{R_{\text{max}}}{2}} \left(R \sqrt{-1 + \frac{R_{\text{max}}}{R}} + \arctan\left(\sqrt{-1 + \frac{R_{\text{max}}}{R}}\right)\right),
	\end{equation}
	where on the horizon $R_+ = 2$, we have $T(R_+) = T(2)$. This equation describes the evolution of the star's radius from an arbitrary initial value $R = R_{\text{max}}>R_+$ to $R = 0$ within a finite time $T(0)$ (see Fig. (\ref{f2}) for $R_{\text{max}} = 5$).
	
	For the apparent horizon, substituting eq. (\ref{sch0}) into eq. (\ref{rapc}) gives
	\begin{equation}\label{rapschc}
		R_{\text{ap}}(R) = \frac{R \sqrt{-1 + \frac{R_{\text{max}}}{2}}}{\sqrt{-1 + \frac{R_{\text{max}}}{R}}}.
	\end{equation}
	Solving for $R$ in terms of $R_{\text{ap}}$ and substituting into eq. (\ref{tschc}), we numerically computed the time evolution of the apparent horizon as a function of radius as $T_{\text{ap}}(R)$ and plotted in Fig. (\ref{f2}) for $R_{\text{max}} = 5$, where $T(R_+) = T_{\text{ap}}(R_+)$. To check whether the apparent horizon is timelike or spacelike, substitute eq. (\ref{sch0}) into eq. (\ref{n2c}). We find:
	\begin{equation}
		y = \frac{3 + R_{\text{max}}(-8 + 3R_{\text{max}})}{2(-1 + R_{\text{max}})^2},
	\end{equation}
	which leads to $n^2 > 0$. Thus, during gravitational collapse, the apparent horizon remains timelike.
	
	For the event horizon's evolution, substituting eq. (\ref{sch0}) into eq. (\ref{reh}) gives:
	\begin{equation}\label{rehschc}
		R_{\text{eh}}(R) = R + 2R \sqrt{-1 + \frac{R_{\text{max}}}{2}} \left(-\arctan\left(\sqrt{-1 + \frac{R_{\text{max}}}{2}}\right) + \arctan\left(\sqrt{-1 + \frac{R_{\text{max}}}{R}}\right)\right),
	\end{equation}
	where the integration constant is determined from $R_{\text{eh}}(R_+) = R_+$. Solving this equation for $R$ in terms of $R_{\text{eh}}$ and substituting into eq. (\ref{tsrc}), we numerically computed $T_{\text{eh}}(R)$, where $T(R_+) = T_{\text{ap}}(R_+)=T_{\text{eh}}$. For $R_{\text{max}} = 5$, we plotted the results in Fig. (\ref{f2}). The event horizon begins growing during the collapse from $R = R_i$ and eventually crosses both the apparent horizon and the star's surface at $R = R_+$. As shown in Fig. (\ref{f2}), the overall behavior of the collapsing star in the $k = 1$ case is very similar to the $k = 0$ case.
	
	In the case where the star starts collapsing from $R_0 < R_{\text{max}}$, substituting eq.~(\ref{sch0}) into eq.~(\ref{tsrc0}) gives the evolution of the star's surface, and apparent and event horizons evolutions are given by (\ref{rehc}) and (\ref{rapc}). For example, for $R_{\text{max}} = 5$ and $R_0 = 4$, we recover the OS collapse, where the total collapse time is shorter. We will discuss in detail the collapse time and event horizon formation time in Section~5.2. In general, the collapse process is the same as discussed in Fig.~\ref{f2}.
	
	As explained earlier, the forms of eqs. (\ref{den}) and (\ref{pre}) remain unchanged for $k = 1$. Thus, the gravitational mass, density, and pressure of the collapsing star remain the same as in the $k = 0$ case. Additionally, the matter field satisfies all energy conditions.
	
	Comparing Figs. (\ref{f1}) and (\ref{f2}) we figure out that: when \( k = 0 \), space is flat, and the collapse happens more quickly. The event horizon forms earlier and grows more slowly, while the trapped surfaces appear sooner. On the other hand, when \( k = 1 \), space is curved. Here, the event horizon forms later but rises more sharply, and the trapped surfaces take longer to appear. This means the collapse starts slower in a closed space compared to an open one. Despite these differences, both cases eventually lead to the formation of a black hole,  but the timing and characteristics of the collapse are shaped by whether the space is open (\( k = 0 \)) or closed (\( k = 1 \)).

\end{appendix}

\bibliographystyle{JHEP}

\end{document}